\documentclass[a4paper,11pt]{article}%

\usepackage{graphicx,amsmath,amsmath,amsfonts,amssymb,dcolumn,amsthm,euscript,hyperref,xcolor}

\usepackage[utf8]{inputenc}
\usepackage[english]{babel}

\begin{document}
	
	\date{}
	\title{\textbf{On the renormalization of a generalized supersymmetric version of the maximal Abelian gauge}}
	\author{\textbf{M.~A.~L.~Capri}$^{a}$\thanks{caprimarcio@gmail.com}\,\,,
		\textbf{R.~C.~Terin}$^{a,b}$\thanks{rodrigoterin3003@gmail.com}\,\,,
				\textbf{H.~C.~Toledo}$^{a}$\thanks{henriqcouto@gmail.com}\,\,\,\\[2mm]
		{\small \textnormal{$^{a}$  \it Departamento de F\'{\i }sica Te\'{o}rica, Instituto de F\'{\i }sica, Universidade do Estado do Rio de Janeiro, UERJ,}}
		\\ \small \textnormal{\phantom{$^{a}$} \it Rua S\~{a}o Francisco Xavier 524, 20550-013, Rio de Janeiro, RJ, Brasil}
        \\[2mm] \small \textnormal{$^{b}$  \it Sorbonne Universit\'{e}, CNRS, Laboratoire de Physique Th\'{e}orique de la Mati\`{e}re Condens\'{e}e},		
		\\ \small \textnormal{\phantom{$^{b}$} \it LPTMC, F-75005 Paris, France}}
	
	\maketitle
	
	\begin{abstract}
    \noindent In this work we present an algebraic proof of the renormazibility of the super-Yang-Mills action quantized in a generalized supersymmetric version of the maximal Abelian gauge. The main point stated here is that the generalized gauge depends on a set of infinity gauge parameters in order to take into account all possible composite operators emerging from the dimensionless character of the vector superfield. At the end, after the removal of all ultraviolet divergences, it is possible to specify values to the gauge parameters in order to return to the original supersymmetric maximal Abelian gauge, first presented in Phys.\ Rev.\ D {\bf 91}, no. 12, 125017 (2015), Ref.~\cite{Capri:2015usa}.
	\end{abstract}
	\baselineskip=13pt
	
	
	
	\maketitle
\section{Introduction}

In the understanding of the quarks confinement mechanism, some formulations of the Yang-Mills theory in specific gauge conditions, like the Landau gauge, the maximal Abelian gauge, the Curci-Ferrari gauge and etc, are explored. In particular, the maximal Abelian gauge permits us to approach the notion of Abelian projection, one of the main ideas regarding quarks confinement \cite{tHooft:1981bkw}. Here, the emergence of the magnetic monopoles, provided by the Abelian degrees, can be understood as a confinement mechanism. However, in the supersymmetric scenario, the subject of confinement leads us directly to the ADS/CFT correspondence where there would be a duality between a ten-dimensional low-temperature string theory with the strong couplings of a $ N = 4 $ theory of super-Yang-Mills in four dimensions \cite{Gubser:2009md}.\\\\In fact, supersymmetric theories at finite temperature can reveal fundamental properties, similar to those of weak interactions in a plasma of quarks and gluons. Thus, 
it may be possible to study the transition from the deconfinement phase in an analogous way to the non-supersymmetric case, i.e. 
to analyse phase transitions due to the emergence of singularities in the Abelian sector of a super-Yang-Mills theory.\\\\In this work, our focus is on the proof of the renormazibility of the super-Yang-Mills theory in the supersymmetric maximal Abelian gauge, as proposed in \cite{Capri:2015usa}. However, since the vector superfield, $V(x,\theta,\bar{\theta})$, is dimensionless\footnote{See Table~\ref{quantum_numbers} and the discussion in the Sect.~\ref{generalized}} the 
construction of the most general counterterm, which cancels all divergences of the theory, becomes complicated because there are infinite insertions depending on the vector superfield. Therefore, the version of the maximal Abelian gauge presented in \cite{Capri:2015usa} is not unique, it is actually the simplest one, and the correct definition of this gauge in superspace seemed to be ambiguous. The solution for this problem is a generalization of the gauge fixing condition in order to include all ambiguities in the definition of the gauge. Namely,
\begin{eqnarray}
\bar{\mathcal{D}}^{2}\mathcal{D}^{2}V^{i} = 0 &\longrightarrow& \bar{\mathcal{D}}^{2}\mathcal{D}^{2}\omega^{i}(V)= 0\,, \\
\bar{\mathcal{D}}^{2}\mathcal{D}^{2}\left(V^{a} - \frac{i}{2}f^{abi}V^{i}V^{b}\right) = 0 &\longrightarrow& \bar{\mathcal{D}}^{2}\mathcal{D}^{2}\omega^{a}(V)= 0\,,
\end{eqnarray}
where the expressions in the l.h.s. are the original conditions of the maximal Abelian gauge in superspace introduced in \cite{Capri:2015usa}, and $\omega^{a}(V)$ and $\omega^{i}(V)$ are general power series in $V$ obeying some symmetry criterions that will be clear later in Sect.~\ref{generalized}. Also, the indices conventions will be clarified in Sect.~\ref{simplest} and in Appendix~\ref{MAGYM}, but at the moment is sufficient to say that the index $i$ is related to the Abelian components of the internal symmetry group, the SU($n$) group, and the indices $\{a,b\}$ to the non-Abelian ones.
One advantage of this method lies in the fact that it defines a general class of non-linear gauges depending of a set of gauge parameters that can be chosen in a suitable way after the renormalization procedure, or, in other word, after the removal of the ultraviolet divergences of the theory. Then, the supersymmetric version of the maximal Abelian gauge is viewed here as a particular case of this general gauge. Actually, even the Landau gauge can be classified into this general class \cite{Capri:2018gpu}.\\\\The paper is organized as follows: The Sect.~\ref{presentation} is subdivied in five subsections in which we made a brief review on the supersymmetric extension of the maximal Abelian gauge, as presented in \cite{Capri:2015usa}, and how it can be generalized. The classical starting point action is also defined in this section and its rich symmetry content is meticulously discussed; In the Sect.~\ref{renorm}, we perform the proof of renormazibility of the theory following the Algebraic renormalization set up outlined in Refs~\cite{Capri:2018gpu,Becchi:1975nq,Piguet:1980nr,Piguet:1995er,Clark:1977pq,Piguet:1981fb,Piguet:1981hh,Piguet:1987pe,Piguet:1996ys}\footnote{The Refs~\cite{Becchi:1975nq,Piguet:1980nr,Piguet:1995er} are standard references on the procedure of the algebraic renormalization, while Refs~\cite{Capri:2018gpu,Clark:1977pq,Piguet:1981fb,Piguet:1981hh,Piguet:1987pe,Piguet:1996ys} relate to the renormalization in the presence of dimensionless vector superfield in the Landau and the linear covariant gauges. The problem of dealing with the renormalization of a theory with dimensionless fields is also approached in Refs~\cite{Capri:2017bfd,Capri:2017npq} in the case of a Stueckelberg-like field. }. The renormalization factors (the ``famous'' $Z$ factors) of fields, sources and parameters of the theory are obtained, becoming clear that the renormalization of the vector superfield is non-linear and its components (the Abelian and non-Abelian ones) are mixed in quantum corrections into a matricial renormalization; In Sect.~\ref{conclusions}, we conclude this work with some final discussions and perspectives for future works. Finally, in Appendix~\ref{MAGYM}, we displayed, for the sake of the reader, a review on the maximal Abelian gauge for ordinary SU($n$) Yang-Mills theory and some relevant notations and conventions are properly defined.

\section{The supersymmetric maximal abelian gauge}
\label{presentation}

\subsection{The simplest formulation}
\label{simplest}
In this section we perform a brief review on the supersymmetric formulation of maximal Abelian gauge, first presented in \cite{Capri:2015usa}\footnote{See also Appendix \ref{MAGYM} for a review of the maximal Abelian gauge for the ordinary Yang-Mills theory in the SU($n$) group.}. The main characteristic of this gauge is to explicitly split the Abelian and non-Abelian sectors of the gauge symmetry group. In the SU($n$) group one can decompose the vector superfield $V(x,\theta,\bar{\theta})$ in terms of the group generators as, 
\begin{equation}
V(x,\theta,\bar{\theta}) = \sum_{A=1}^{n^{2}-1}V^{A}(x,\theta,\bar{\theta})T^{A} = \sum_{a=1}^{n(n-1)}V^{a}(x,\theta,\bar{\theta})T^{a}+\sum_{i=1}^{n-1} V^{i}(x,\theta,\bar{\theta})T^{i}\,,\\
\end{equation}
where $T^{A}$ stands by the $n^{2}-1$ group generators of SU($n$) which can be split in the $n(n-1)$  off-diagonal generators, $T^{a}$, and in the $n-1$ diagonal generators, $T^{i}$, which form an Abelian subgroup of SU($n$) also known as the Cartan subgroup. Also, we have adopted here capital letters $\{A,B,C,\dots\}$, running from 1 to $n^{2}-1$, for the full SU($n$) group; the labels $\{a,b,c,d,e,...\}$, running from 1 to $n(n-1)$, for the off-diagonal sector; and the indices $\{i,j,k,l,\dots\}$, running from 1 to $n-1$, for the Abelian sector. Of course, we will assume, from now on, the Einstein summation convention for repeated indices\footnote{The algebra of the generators and some useful identities can be found in Appendix \ref{MAGYM}.}.\\\\In oder to make clear our notations and conventions, let us display here the vector superfield in terms of its components: 
\begin{eqnarray}
V^{A} & = & C^{A}(x)+\theta^{\alpha}\chi^{A}_{\alpha}(x)+\bar{\theta}_{\dot{\alpha}}\bar{\chi}^{A\dot{\alpha}}(x)+\frac{1}{2}\theta^{2}M^{A}(x)+\frac{1}{2}\bar{\theta}^{2}\bar{M}^{A}(x)+2\theta^{\alpha}\sigma_{\alpha\dot{\alpha}}^{\mu}\bar{\theta}^{\dot{\alpha}}A^{A}_{\mu}(x)\nonumber\\
&+&\frac{1}{2}\bar{\theta}^{2}\theta^{\alpha}\lambda^{A}_{\alpha}(x)+\frac{1}{2}\theta^{2}\bar{\theta}_{\dot{\alpha}}\bar{\lambda}^{A\dot{\alpha}}(x)+\frac{1}{4}\theta^{2}\bar{\theta}^{2}\mathfrak{D}^{A}(x)\,,
\label{V}
\end{eqnarray}
where, $\theta^{\alpha}$ $(\alpha=1,2)$ and $\bar\theta_{\dot{\alpha}}$ $(\dot{\alpha}=\dot{1},\dot{2})$ are the fermionic coordinates of the superspace; and  $\{C,\chi_{{\alpha}},\bar\chi_{\dot{\alpha}},M,\bar{M},A_{\mu},\lambda_{\alpha},\bar\lambda_{\dot\alpha},\mathfrak{D}\}$ the superfield components in the adjoint representation of SU($n$) group\footnote{We also have in Eq.~\eqref{V} that $\theta^{2}=\theta^{\alpha}\theta_{\alpha}$, $\bar\theta^{2}=\bar{\theta}_{\dot{\alpha}}\bar{\theta}^{\dot{\alpha}}$ and $\sigma^{\mu}=(\mathbf{1},\sigma_1,\sigma_2,\sigma_3)$, being $\mathbf{1}$ the identity matrix of order two and $(\sigma_1,\sigma_2,\sigma_3)$ the usual Pauli matrices.}.\\\\It is well known that in the quantization of a gauge field theory (being supersymmetric or not) an additional condition (or a constraint, or a gauge-fixing condition) for the gauge field needs to be implemented. Following \cite{Capri:2015usa}, such additional condition can be chosen as       
\begin{eqnarray}
\bar{\mathcal{D}}^{2}\mathcal{D}^{2}\left(V^{a} - \frac{i}{2}f^{abi}V^{i}V^{b}\right) &=& 0\,,
\label{SUPERMAG2}\\
\bar{\mathcal{D}}^{2}\mathcal{D}^{2}V^{i} &=& 0 \,,\label{SUPERMAG1}
\end{eqnarray}
where, $\mathcal{D}_{\alpha}$ and $\bar{\mathcal{D}}_{\dot{\alpha}}$ are, respectively, the chiral and antichiral covariant derivatives, given by 
\begin{eqnarray}
\mathcal{D}_{\alpha} &=& \frac{\partial}{\partial\theta^{\alpha}} - i\sigma^{\mu}_{\alpha \dot{\alpha}}\bar{\theta}^{\dot{\alpha}}\partial_{\mu},\\
\bar{\mathcal{D}}_{\dot{\alpha}} &=& -\frac{\partial}{\partial\bar{\theta}^{\dot{\alpha}}} - i\theta^{\alpha}\sigma^{\mu}_{\alpha \dot{\alpha}}\partial_{\mu}\,,
\end{eqnarray}
and $f^{abi}$ represents one of the possible types of structure constants appearing in the algebra associated with the SU($n$) group, see details in Appendix \ref{MAGYM}. Considering the constraints \eqref{SUPERMAG2} and \eqref{SUPERMAG1}  written in terms of the components of the superfield and, together with the so-called Wess-Zumino gauge, i.e. taking the lower components $\{C,\chi,\bar{\chi},M,\bar{M}\}$ equal to zero, it is possible to obtain the usual maximal Abelian gauge \cite{Capri:2015usa}:  
\begin{eqnarray}
\partial_{\mu}A^{a\mu} &=& f^{abi}A^{i}_{\mu}A^{b\mu}\,,\\
\partial_{\mu}A^{i\mu} &=& 0\,.
\end{eqnarray}
As we shall see later, the superfield $V(x,\theta,\bar{\theta})$ is dimensionless. Therefore, the off-diagonal gauge-fixing condition \eqref{SUPERMAG2} is the simplest way to define the supersymmetric version of the maximal Abelian gauge, while the diagonal condition \eqref{SUPERMAG1} corresponds to a Landau-like gauge. Thus, we will name the gauge defined by Eqs.~\eqref{SUPERMAG2} and \eqref{SUPERMAG1} as SSMAG meaning ``Simplest-Super-Maximal-Abelian-Gauge''.  In Section \ref{generalized}, we will discuss a generalized version of SSMAG that we will call GSMAG meaning ``Generalized-Super-Maximal-Abelian-Gauge''.

\subsection{The Faddeev-Popov quantization and BSRT symmetry}
Since we have defined the gauge conditions of the SSMAG, Eqs \eqref{SUPERMAG2} and \eqref{SUPERMAG1}, we are able to carry out the Faddeev-Popov quantization procedure \cite{Piguet:1996ys,Gates:1983nr,Bailin:1986wt,Ali:1994wm}. First of all, let us start by the super-Yang-Mills (SYM) action: 
\begin{equation}
S_{\mathrm{SYM}}=-\frac{1}{64g^{2}}\,\mathrm{tr}\int d^{4}xd^{2}\theta\,\mathcal{W}^{\alpha}\mathcal{W}_{\alpha}+\mathrm{c.c.}
=-\frac{1}{128g^{2}}\int d^{4}xd^{2}\theta\,W^{A\alpha}W^{A}_{\alpha}+\mathrm{c.c.}\,,
\label{SYM_action}
\end{equation}
where, $g$ is the coupling constant and $\mathcal{W}_{\alpha}$ the chiral-field strength, given by
\begin{equation}
\mathcal{W}_{\alpha}=W^{A}_{\alpha}T^{A}=\bar{\mathcal{D}}^{2}(e^{-V}\mathcal{D}_{\alpha}e^{V})\,.
\end{equation}
It can be checked that the action \eqref{SYM_action} is invariant by the following infinitesimal gauge transformations,
\begin{eqnarray}
V &\to&\ V'=V+\delta V\,,\nonumber\\
\delta V &=& \frac{i}{2}L_{V}(\Lambda+\bar\Lambda)+\frac{i}{2}\left[L_{V}\,\mathrm{coth}\left(\frac{1}{2}L_{V}\right)\right]
(\Lambda-\bar\Lambda)\nonumber\\
&=&i(\Lambda-\bar\Lambda)
+\frac{1}{2}[V,\Lambda+\bar\Lambda]+\frac{i}{12}[V,[V,\Lambda-\bar\Lambda]]+\mathcal{O}(V^{3})\,,
\label{gauge_transf}
\end{eqnarray}
where $L_{V}\bullet=[V,\bullet]$ and $\Lambda=\Lambda^{A}T^{A}$ are chiral infinitesimal superfields, while $\bar\Lambda=\bar\Lambda^{A}T^{A}$ are antichiral superfields. As a gauge theory, the correct quantization\footnote{Actually, the correct quantization of a gauge theory is an open problem until now. The Faddeev-Popov method is considered to be correct only at perturbative level, but, at non-perturbative level, other effects, as the Gribov ambiguity problem, show up and have to be taken into account. In this work we will restrict ourselves to the Faddeev-Popov quantization method.} needs the implementation of a gauge-fixing condition. The Faddeev-Popov method correponds to a way of introducing a constraint in the functional integral for a gauge theory. In this method the SYM action needs to be supplemented, in the Feynman path integrals, by a term including such constraint. Then, according to the Faddeev-Popov method, the SYM action is replaced by 
\begin{equation}
S_{\mathrm{FP}}=S_{\mathrm{SYM}}+S_{\mathrm{gf}}\,,
\label{FP}
\end{equation}
where $S_{\mathrm{gf}}$ is the gauge-fixing action. In our case, the gauge-fixing action for the SSMAG is, following \cite{Capri:2015usa}, given by
\begin{eqnarray}
S_{\mathrm{gf}}=\frac{1}{8}\int d^{4}xd^{2}\theta\left[B^{a}\bar{\mathcal{D}}^{2}\mathcal{D}^{2}\left(V^{a}-\frac{i}{2}
f^{abi}V^{i}V^{b}\right)+B^{i}\bar{\mathcal{D}}^{2}\mathcal{D}^{2}V^{i}\right]+\mbox{``ghost terms''}+\mathrm{c.c.}\,.
\label{gf_0}
\end{eqnarray} 
The action above needs several explanations. First, the fields $B^{a}$ and $B^{i}$ play the role of Lagrange multipliers enforcing the SSMAG conditions for the off-diagonal and diagonal sectors, respectively. It is immediately checked that their classical equations of motion coincide with the constraints \eqref{SUPERMAG2} and \eqref{SUPERMAG1}. Also, it is easy to notice that they are chiral superfields. Then, in the complex conjugated part, standing by ``c.c.'', the antichiral superfields $\bar{B}^{a}$ and $\bar{B}^{i}$ must appear.  The term ``ghost terms'' represents here all terms involving the well-known Faddeev-Popov ghost fields. In order to avoid any confusion, before presenting this term explicitly let us display here the four sets of ghosts that we have to deal: 
\begin{itemize}
\item{The off-diagonal chiral ghosts: $\{c^{a},c^{a}_{\star}\};$}
\item{The off-diagonal antichiral ghosts: $\{\bar{c}^{a},\bar{c}^{a}_{\star}\};$}
\item{The diagonal chiral ghosts: $\{c^{i},c^{i}_{\star}\};$}
\item{The diagonal antichiral ghosts: $\{\bar{c}^{a},\bar{c}^{a}_{\star}\}.$}
\end{itemize}
In each set listed above the subscribed symbol ``$\,\star\,$'' indicates the antighost and the bar indicates the { antichiral} character of the field. For example, $c^{a}$ is the {\it off-diagonal chiral ghost}, $c^{i}_{\star}$ is the {\it diagonal chiral antighost} and $\bar{c}^{a}_{\star}$ is the {\it off-diagonal { antichiral} antighost}. Another point that must be explained here is that, although the gauge invariance has been lost in the gauge-fixing procedure, the Faddev-Popov action $S_{\mathrm{FP}}$, is left invariant by a set of transformations, the so-called BRST tranformations, listed bellow: 
\begin{itemize}
\item{Transformations of the components of the vector superfield $V(x,\theta,\bar\theta)$:
\begin{eqnarray}
sV^{a}&=&i(c^{a}-\bar{c}^{a})
-\frac{1}{2}f^{abc}V^{b}(c^{c}+\bar{c}^{c})
-\frac{1}{2}f^{abi}V^{b}(c^{i}+\bar{c}^{i})
+\frac{1}{2}f^{abc}V^{i}(c^{b}+\bar{c}^{b})
+\mathcal{O}(V^{2})\,,\nonumber\\
sV^{i}&=&i(c^{i}-\bar{c}^{i})
-\frac{1}{2}f^{abi}V^{a}(c^{b}+\bar{c}^{b})
+\mathcal{O}(V^{2})\,;
\label{brst1}
\end{eqnarray}}
\item{Transformations of the components of chiral superfields $\{c,c_{\star},B\}$:
\begin{equation}
\begin{tabular}{lll}
$\displaystyle sc^{a}=f^{abi}c^{b}c^{i}+\frac{1}{2}f^{abc}c^{b}c^{c}\,,\qquad$
&$\displaystyle sc^{i}=\frac{1}{2}f^{abi}c^{a}c^{b}\,,$&$\!\phantom{\Big|}$\cr
$sc^{a}_{\star}=B^{a}\,,$&$sc^{i}_{\star}=B^{i}
\,,$&$\!\phantom{\Big|}$\cr
$sB^{a}=0\,,$&$sB^{i}=0\,;$&$\!\phantom{\Big|}$\cr
\end{tabular}
\label{brst2}
\end{equation}}
\item{Transformations of the components of the { antichiral} superfields $\{\bar{c},\bar{c}_{\star},\bar{B}\}$:
\begin{equation}
\begin{tabular}{lll}
$\displaystyle s\bar{c}^{a}=f^{abi}\bar{c}^{b}\bar{c}^{i}+\frac{1}{2}f^{abc}\bar{c}^{b}\bar{c}^{c}\,,\qquad$
&$\displaystyle s\bar{c}^{i}=\frac{1}{2}f^{abi}\bar{c}^{a}\bar{c}^{b}\,,$&$\!\phantom{\Big|}$\cr
$s\bar{c}^{a}_{\star}=\bar{B}^{a}\,,$&$s\bar{c}^{i}_{\star}=\bar{B}^{i}
\,,$&$\!\phantom{\Big|}$\cr
$s\bar{B}^{a}=0\,,$&$s\bar{B}^{i}=0\,.$&$\!\phantom{\Big|}$\cr
\end{tabular}
\label{brst3}
\end{equation}}
\end{itemize}
As one can see the BRST transformations of the vector superfield is similar to the infinitesimal gauge tranformations \eqref{gauge_transf}, just replacing the infinitesimal gauge parameters $\{\Lambda,\bar{\Lambda}\}$ by the ghosts $\{c,\bar{c}\}$. Therefore, the SYM action is automatically invariant by the set of transformations \eqref{brst1}. Also, the BRST operator, $s$, is nilpotent, i.e. $s^{2}=0$ and, thanks to this remarkable property, one can finally write the gauge-fixing term as a full BRST variation:
\begin{eqnarray}
S_{\mathrm{gf}}&=&\frac{1}{8}s\int dV\bigg[c^{a}_{\star}\mathcal{D}^{2}\left(V^{a}-\frac{i}{2}f^{abi}V^{i}V^{b}\right)
+c^{i}_{\star}\mathcal{D}^{2}V^{i}
+\bar{c}^{a}_{\star}\bar{\mathcal{D}}^{2}\left(V^{a}+\frac{i}{2}f^{abi}V^{i}V^{b}\right)
+\bar{c}^{i}_{\star}\bar{\mathcal{D}}^{2}V^{i}
\bigg]\nonumber\\
&=&\frac{1}{8}\int dV\bigg[B^{a}\mathcal{D}^{2}\left(V^{a}-\frac{i}{2}f^{abi}V^{i}V^{b}\right)
+B^{i}\mathcal{D}^{2}V^{i}
+\bar{B}^{a}\bar{\mathcal{D}}^{2}\left(V^{a}+\frac{i}{2}f^{abi}V^{i}V^{b}\right)
\nonumber\\
&&
+\bar{B}^{i}\bar{\mathcal{D}}^{2}V^{i}\bigg]
-\frac{1}{8}\int dV\bigg[c^{a}_{\star}\mathcal{D}^{2}s\left(V^{a}-\frac{i}{2}f^{abi}V^{i}V^{b}\right)
+c^{i}_{\star}\mathcal{D}^{2}sV^{i}
\nonumber\\
&&
+\bar{c}^{a}_{\star}\bar{\mathcal{D}}^{2}s\left(V^{a}+\frac{i}{2}f^{abi}V^{i}V^{b}\right)
+\bar{c}^{i}_{\star}\bar{\mathcal{D}}^{2}sV^{i}\bigg]\,,
\label{SYM+MAG}
\end{eqnarray}
where $dV\equiv d^{4}xd^{2}\theta d^{2}\bar{\theta}$ is the superspace element volume. The action above is evidently invariant due to the nilpotency property and thus the Faddeev-Popov action \eqref{FP} is BRST invariant. Summarizing our current situation, we have at our disposal the BRST invariant action \eqref{FP}, representing the $N=1$ SYM theory for SU($n$) group quantized in the so-called SSMAG. The next step would be the study of the renormalizability of this model. In order to achieve this aim let us first study the symmetry content and the Ward identities of the model. 

\subsection{Local composite operator formalism and the Ward identities}
In this section we would like to study the symmetry content of the action \eqref{FP}. We already known that action \eqref{FP} is BRST invariant and the BRST tranformations are nonlinear. In order to deal with such nonlinear symmetry, and other possible {nonlinear} identities of the model, we need to make use of the local composite operator formalism \cite{Piguet:1995er}. For this purpose, let us consider the following action:
\begin{equation}
S=S_{\mathrm{FP}}+S_{\mathrm{ext}}\,.\label{S}
\end{equation}
The external action $S_{\mathrm{ext}}$ is a term depending on external sources coupled to some local composite operators. More specifically, we have
\begin{eqnarray}
S_{\mathrm{ext}}[\Omega,L,\bar{L},R,P]&=&\int dV \,\Big[\Omega^{a}\left(sV^{a}\right)+\Omega^{i}\left(sV^{i}\right)\Big]
+\int d^{4}x d^{2}\theta\,\Big[L^{a}\left(sc^{a}\right)+L^{i}\left(sc^{i}\right)\Big]
\nonumber\\
&&+\int d^{4}x d^{2}\bar\theta\,\Big[\bar{L}^{a}\left(s\bar{c}^{a}\right)+\bar{L}^{i}\left(s\bar{c}^{i}\right)\Big]
+\int dV\, P^{a}\Big(f^{abi}V^{i}V^{b}\Big)\nonumber\\
&&-\int dV\, R^{a}\,s\Big(f^{abi}V^{i}V^{b}\Big)
\,,
\label{external_1}
\end{eqnarray} 
where the BRST invariance of the external term is guaranteed by the BRST transformations of the external sources bellow:
\begin{eqnarray}
&s\Omega^{a}=0\,,\qquad s\Omega^{i}=0\,,\qquad sL^{a}=0\,,\qquad s\bar{L}^{a}=0\,,\qquad sL^{i}=0\,,\qquad s\bar{L}^{i}=0\,,&\nonumber\\
&sR^{a}=P^{a}\,,\qquad sP^{a}=0\,.&
\end{eqnarray}
Notice that the sources $\{\Omega,L,\bar{L}\}$ are coupled to the {nonlinear} BRST transformations, while the sources $P^{a}$  and $R^{a}$ are coupled to the composite operator $f^{abi}V^{i}V^{b}$ and its BRST transformation, respectively. Furthermore, the sources $\{R,P\}$ form the so-called BRST doublet and the last two terms of \eqref{external_1} can be written as an exact BRST variation. In fact, the external term can be completely written as a BRST variation:
\begin{eqnarray}
S_{\mathrm{ext}}[\Omega,L,\bar{L},R,P]&=&s\,\bigg[\,\int dV \,(-\Omega^{a}V^{a}-\Omega^{i}V^{i})
+\int d^{4}x d^{2}\theta\,(L^{a}c^{a}+L^{i}c^{i})
\nonumber\\
&&+\int d^{4}x d^{2}\bar\theta\,(\bar{L}^{a}\bar{c}^{a}+\bar{L}^{i}\bar{c}^{i})
+\int dV\, f^{abi}R^{a}V^{i}V^{b}\,\bigg]
\,.
\label{external_2}
\end{eqnarray}     
The external sources are introduced here as a mathematical tool that allow us to define some important {Green} functions of the model and to write in a well-defined manner the {nonlinear} Ward identities of the model. As these sources vanish the action $S$, Eq.~\eqref{S}, coincides with $S_{\mathrm{FP}}$, Eq.~\eqref{FP}. Therefore, action \eqref{S} is suitable to study the symmetry content in terms of Ward identities, including the {nonlinear} ones.  As a last remark before discussing the Ward identities, we would like to call attention to the fact that the BRST operator, $s$, as well as the Faddeev-Popov ghosts are Grassmann variables and carry a quantum number named {\it ghost number} $(g\#)$. In Table \ref{quantum_numbers} we displayed the quantum numbers of the fields and sources of the theory, including the mass dimensions. Notice that the sources with odd ghost number, as $\Omega$ and $R$, are anticommuting, while the sources with even ghost number, as $L$, $\bar{L}$ and $P$, are commuting\footnote{In general, if the combination $2d+(g\#)$, with $d$ being the mass dimension, is an even number, the corresponding object (a field, a source, a paremeter or an operator) is commuting. Otherwise it is anticommuting.}.\\\\Now we are able to present the set of Ward identities enjoyed by the action \eqref{S}. These identities represent the set of all symmetries of the theory being fundamental for the proof of its renormalizability. 

\subsubsection{The Slavnov-Taylor identity} 
The BRST symmetry can be written as a functional identity as follows
\begin{eqnarray}
\mathcal{B}(S)&:=&\int dV\,\left(
\frac{\delta S}{\delta\Omega^{a}}\frac{\delta S}{\delta V^{a}}
+\frac{\delta S}{\delta\Omega^{i}}\frac{\delta S}{\delta V^{i}}
+P^{a}\frac{\delta S}{\delta R^{a}}\right)\nonumber\\
&&+\int d^{4}x d^{2}\theta\,\left(\frac{\delta S}{\delta L^{a}}\frac{\delta S}{\delta c^{a}}
+\frac{\delta S}{\delta L^{i}}\frac{\delta S}{\delta c^{i}}
+B^{a}\frac{\delta S}{\delta c^{a}_{\star}}
+B^{i}\frac{\delta S}{\delta c^{i}_{\star}}\right)\nonumber\\
&&
+\int d^{4}xd^{2}\bar\theta\,\left(\frac{\delta S}{\delta \bar{L}^{a}}\frac{\delta S}{\delta \bar{c}^{a}}
+\frac{\delta\Sigma}{\delta \bar{L}^{i}}\frac{\delta S}{\delta \bar{c}^{i}}
+\bar{B}^{a}\frac{\delta S}{\delta \bar{c}^{a}_{\star}}
+\bar{B}^{i}\frac{\delta S}{\delta \bar{c}^{i}_{\star}}\right)\nonumber\\
&=&0\,.
\label{ST}
\end{eqnarray}

\subsubsection{The diagonal gauge-fixing equations}
The classical equations of motion of the diagonal Lagrange multipliers $B^{i}$ and $\bar{B}^{i}$, being linear in the fields,  can be recognized as valid equations of the so-called Quantum Action Principle (QAP) \cite{Piguet:1995er}. Namely,
\begin{equation}
\frac{\delta S}{\delta B^{i}}=\frac{1}{8}\,\bar{\mathcal{D}}^{2}\mathcal{D}^{2}V^{i}\,,\qquad
\frac{\delta S}{\delta \bar{B}^{i}}=\frac{1}{8}\,\mathcal{D}^{2}\bar{\mathcal{D}}^{2}V^{i}\,.
\label{diag_gaugefixing}
\end{equation}
\subsubsection{The off-diagonal gauge-fixing equations}
In contrast with the diagonal equations of motion of the Lagrange multipliers $\{B^{i},\bar{B}^{i}\}$, the equations of motion of the off-diagonal Lagrange multipliers $\{B^{a},\bar{B}^{a}\}$ are {nonlinear}. It is a direct consequence of the nonlinearity of the SSMAG. However, with the help of the insertion of the local composite operator $f^{abi}V^{i}V^{b}$ we can write the equations of motion of the fields $\{B^{a},\bar{B}^{a}\}$ as the following functional identities:
\begin{eqnarray}
\frac{\delta S}{\delta B^{a}}+\frac{i}{16}\,{\bar{\mathcal{D}}}^{2}\mathcal{D}^{2}\frac{\delta S}{\delta P^{a}}&=&
\frac{1}{8}\,{\bar{\mathcal{D}}}^{2}\mathcal{D}^{2} V^{a}\,,\label{chiral_off_diag_gaugefixing_eq}\\
\nonumber\\
\frac{\delta S}{\delta \bar{B}^{a}}-\frac{i}{16}\,\mathcal{D}^{2}{\bar{\mathcal{D}}}^{2}\frac{\delta S}{\delta P^{a}}&=&
\frac{1}{8}\,\mathcal{D}^{2}{\bar{\mathcal{D}}}^{2} V^{a}\,.\label{antichiral_off_diag_gaugefixing_eq}
\end{eqnarray}
\subsubsection{The diagonal antighost equations}
From the diagonal antighost equations the following identities can be obtained\footnote{Some authors call these equations as ghost equations. In our nomenclature, however, we have decided to name the equations obtained from the functional derivatives of the antighost fields (chiral or {antichiral}) as {\it antighost equations} and the identities obtained from the functional derivatives of the ghost fields as {\it ghost equations}.}:
\begin{equation}
\frac{\delta S}{\delta c^{i}_{\star}}+\frac{1}{8}\bar{\mathcal{D}}^{2}\mathcal{D}^{2}\frac{\delta S}{\delta\Omega^{i}}=0\,,\qquad
\frac{\delta S}{\delta \bar{c}^{i}_{\star}}+\frac{1}{8}{\mathcal{D}}^{2}\bar{\mathcal{D}}^{2}\frac{\delta S}{\delta\Omega^{i}}=0\,,
\label{diag_anti_ghost_eqs}
\end{equation}
\subsubsection{The off-diagonal antighost equations}
\begin{eqnarray}
\frac{\delta S}{\delta c^{a}_{\star}}+\frac{1}{8}\,\bar{\mathcal{D}}^{2}\mathcal{D}^{2}\frac{\delta S}{\delta \Omega^{a}}
+\frac{i}{16}\,\bar{\mathcal{D}}^{2}\mathcal{D}^{2}\frac{\delta S}{\delta R^{a}}&=&0\,,\label{chiral_off_diag_antighost_eq}\\
\nonumber\\
\frac{\delta S}{\delta \bar{c}^{a}_{\star}}+\frac{1}{8}\,\mathcal{D}^{2}\bar{\mathcal{D}}^{2}\frac{\delta S}{\delta \Omega^{a}}
-\frac{i}{16}\,\mathcal{D}^{2}\bar{\mathcal{D}}^{2}\frac{\delta S}{\delta R^{a}}&=&0\,.\label{antichiral_off_diag_antighost_eq}
\end{eqnarray}
Like the off-diagonal gauge-fixing equations, these identities are only possible in the presence of the composite operator $f^{abi}V^{i}V^{b}$ and its BRST variation. It is important to remark here that in the non-supersymmetric version of the MAG, see Appendix \ref{MAGYM}, the analogues identities cannot be recovered by introducing any composite operators\footnote{Actually they can be recovered but they are completely innocuous.}$^{,}$\footnote{A detailed discussion on the  Ward
identities in the ordinary maximal Abelian gauge can be found in \cite{Fazio:2001rm}.}. Therefore, it seems to be a property of the superspace formulation. 

\subsubsection{The $\mathcal{R}$-invariance}
\begin{eqnarray}
\mathcal{R}(S)&:=&
\sum_{X\in\{V,\Omega,R,P\}}\int dV\,(\delta_{\mathcal{R}} X)\frac{\delta S}{\delta X}\nonumber\\
&&
+\sum_{Y\in\{c,c_\star,B,L\}}\int d^{4}x d^{2}\theta\,
(\delta_{\mathcal{R}} Y)\frac{\delta S}{\delta Y}\,\,\,\,\,
+\sum_{\bar{Y}\in\{\bar{c},\bar{c}_{\star},\bar{B},\bar{L}\}}\int d^{4}x d^{2}\bar\theta\,
(\delta_{\mathcal{R}} \bar{Y})\frac{\delta S}{\delta \bar{Y}}\nonumber\\
&=&0\,.
\end{eqnarray}
The $\mathcal{R}$-variations $(\delta_{\mathcal{R}}Z)$, with $Z$ being any superfield of the theory, are given by
\begin{equation}
\delta_{\mathcal{R}}Z=i\left(n_{Z}+\theta^{\alpha}\frac{\partial}{\partial\theta^{\alpha}}
-\bar{\theta}^{\dot\alpha}\frac{\partial}{\partial\bar{\theta}^{\dot\alpha}}\right)Z\,,
\label{R_var}
\end{equation}
with $n_{Z}$ being the so-called ``$\mathcal{R}$-weight" of the respective superfield $Z$. The $\mathcal{R}$-weight of all objects present in the theory (fields, sources, covariant derivatives and etc.) are displayed in Table \ref{quantum_numbers}.
\subsubsection{The diagonal rigid invariance}
\begin{eqnarray}
\mathcal{W}^{i}(S)&:=&
\sum_{X\in\{V,\Omega,R,P\}}\int dV\,f^{abi}X^{a}\frac{\delta S}{\delta X^{b}}\nonumber\\
&&
+\sum_{Y\in\{c,c_\star,B,L\}}\int d^{4}x d^{2}\theta\,
f^{abi}Y^{a}\frac{\delta S}{\delta Y^{b}}\,\,\,\,\,
+\sum_{\bar{Y}\in\{\bar{c},\bar{c}_{\star},\bar{B},\bar{L}\}}\int d^{4}x d^{2}\bar\theta\,
f^{abi}\bar{Y}^{a}\frac{\delta S}{\delta \bar{Y}^{b}}\nonumber\\
&=&0\,.
\label{rigid_symm}
\end{eqnarray}
Notice that the diagonal rigid symmetry involves transformations only in the off-diagonal components of fields and sources. Also, this symmetry is a consequence of the split of the diagonal and off-diagonal components of the group, which is the main characteristic of the SSMAG, corresponding to a residual U(1)$^{n-1}$ invariance, e.g. see \cite{Dudal:2004rx} for a non-supersymmetric case. In contrast, in the Landau gauge, the rigid symmetry extends to the whole group \cite{Capri:2018gpu,Piguet:1996ys}.

\subsubsection{The diagonal ghost equation}
Another important symmetry for the renormalization procedure is the diagonal ghost equation, given by
\begin{eqnarray}
\mathcal{G}(S) &:=&\int d^{4}xd^{2}\theta\,\left(\frac{\delta S}{\delta c^{i}}+f^{abi}c^{a}_{\star}\frac{\delta S}{\delta B^{b}}\right) + \int d^{4}xd^{2}\bar{\theta}\,\left(\frac{\delta S}{\delta \bar{c}^{i}}+f^{abi}\bar{c}^{a}_{\star}\frac{\delta S}{\delta \bar{B}^{b}}\right)+\int dV\,f^{abi}R^{a}\frac{\delta S}{\delta P^{b}}\nonumber\\
&=&\int d^{4}xd^{2}\theta\,f^{abi}L^{a}c^{b}
+\int d^{4}xd^{2}\bar{\theta}\,f^{abi}\bar{L}^{a}\bar{c}^{b}-\int dV\, f^{abi}\Omega^{a}V^{b}\,.
\label{gh_eq}
\end{eqnarray}
It is important to point out here that the diagonal ghost equation can only be obtained by combining the chiral and  antichiral ghosts, i.e. there is no chiral-ghost equation nor antichiral ghost equation independently. This result was already known in the case of full Landau gauge in \cite{Piguet:1995zz}. Also, the existence of such identity is a signature of the Landau-type diagonal gauge-fixing condition, Eq.~\eqref{SUPERMAG1}.

\begin{table}
\begin{center}
\begin{tabular}{c|c|c|c|c|c|c|c|c|c|c|c|c|c|c|c|c|c}
\hline\hline
&$V$&$c$&$c_{\star}$&$\bar{c}$&$\bar{c}_{\star}$&$B$&$\bar{B}$&$\Omega$&$L$&$\bar{L}$&$\mathcal{D}$&$\bar{\mathcal{D}}$&$\theta$
&$\bar{\theta}\!\phantom{\Big|}$&$R$&$P$&$s$\cr
\hline
$d$&0&0&1&0&1&1&1&2&3&3&$1/2$&$1/2$&$-1/2$&$-1/2$&$2$&$2$&$0$\cr
$g\#$&0&1&$-1$&1&$-1$&$0$&$0$&$-1$&$-2$&$-2$&0&0&0&0&$-1$&$0$&$1$\cr
$n$&0&0&2&0&$-2$&2&$-2$&0&2&$-2$&$-1$&$1$&1&$-1$&$0$&$0$&$0$\cr
\hline\hline
\end{tabular}
\end{center}
\caption{Mass dimension $d$, ghost number $g\#$ and $\mathcal{R}$-weights $n$ of the fields, sources, covariant derivatives and etc.}
\label{quantum_numbers}
\end{table}

\subsection{The generalized formulation}
\label{generalized}

Noticing that the vector superfield $V$ is dimensionless, the SSMAG, given by Eqs \eqref{SUPERMAG2} and \eqref{SUPERMAG1}, could be written, equally well, as
\begin{eqnarray}
\bar{\mathcal{D}}^{2}\mathcal{D}^{2}\bigg(V^{a} - \frac{i}{2}f^{abi}V^{i}V^{b}+\lambda^{aABC}\,V^{A}V^{B}V^{C}+O(V^{4})\bigg) &=& 0\,,
\label{SUPERMAG3}\\
\bar{\mathcal{D}}^{2}\mathcal{D}^{2}\bigg(V^{i}+\eta^{iABC}\,V^{A}V^{B}V^{C}+O(V^{4})\bigg) &=& 0 \,,\label{SUPERMAG4}
\end{eqnarray}
where,
\begin{eqnarray}
\lambda^{aABC}&\in & \Big\{\lambda^{abcd},\lambda^{abci},\lambda^{abij},\lambda^{aijk}\Big\}\,, \\
\eta^{iABC}&\in & \Big\{\eta^{iabc},\eta^{ijab},\eta^{ijka},\eta^{ijkl}\Big\}\,, 
\end{eqnarray}
are invariant tensors constrained by the diagonal rigid invariance \eqref{rigid_symm}. In fact, these tensors are particular linear combinations of the rank-2 invariant tensors $\delta^{ab}$, $\delta^{ij}$ and the structure constants $f^{abc}$ and $f^{abi}$. The coefficients of such linear combinations are gauge parameters that might be taken to zero after the renormalization procedure, recovering then the original SSMAG. Also, it is necessary to remark that the set of Ward identities previously presented, Eqs (\ref{ST}--\ref{gh_eq}), does not prevent us of the redefinitions \eqref{SUPERMAG3} and \eqref{SUPERMAG4}.\\\\Then, due to the above mentioned ambiguity, it is necessary to redefine the gauge-fixing conditions \eqref{SUPERMAG2} and \eqref{SUPERMAG1} by the following generalized versions:
\begin{eqnarray}
\bar{\mathcal{D}}^{2}\mathcal{D}^{2}\omega^{a}(V) & = & 0\,,\label{GSMAG_off}\\\cr
\bar{\mathcal{D}}^{2}\mathcal{D}^{2}\omega^{i}(V) & = & 0\,,\label{GSMAG_diag}
\end{eqnarray}
where,
\begin{eqnarray}
\omega^{a}(V) & = & V^{a}+ \lambda\,f^{abi}V^{i}V^{b}+\lambda^{aABC}\,V^{A}V^{B}V^{C}+\lambda^{aABCD}\,V^{A}V^{B}V^{C}V^{D}+\dots\,,\label{omega_off}\\\cr
\omega^{i}(V) & = & V^{i}+\eta^{iABC}\,V^{A}V^{B}V^{C}+\eta^{iABCD}\,V^{A}V^{B}V^{C}V^{D}+\dots\,.\label{omega_diag}
\end{eqnarray}
The new gauge-fixing conditions \eqref{GSMAG_off} and \eqref{GSMAG_diag} are then a generalized supersymmetric version of the maximal Abelian gauge, taking into account the ambiguities arising from the absence of dimensionality of the vector superfield, and will be refered as the ``Generalized-Super-Maximal-Abelian-Gauge'' (GSMAG), as already mentioned in Section \ref{simplest}. According to the diagonal rigid symmetry \eqref{rigid_symm}, the $\lambda$'s and $\eta$'s tensors must obey the generalized Jacobi identities:
\begin{eqnarray}
0&=&f^{abi}\lambda^{bcde}+f^{cbi}\lambda^{abde}+f^{dbi}\lambda^{acbe}+f^{ebi}\lambda^{acdb}\,,\label{jacobi_lambda}\\
0&=&f^{abi}\lambda^{bcdj}+f^{cbi}\lambda^{abdj}+f^{dbi}\lambda^{acbj}\,,\\
0&=&f^{abi}\lambda^{bcjk}+f^{cbi}\lambda^{abjk}\,,\\
0&=&f^{abi}\lambda^{bijk}\,,\\
\cr
0&=&f^{abi}\lambda^{bcdef}+f^{cbi}\lambda^{abdef}+f^{dbi}\lambda^{acbef}+f^{ebi}\lambda^{acdbf}+f^{fbi}\lambda^{acdeb}\,,\\
0&=&f^{abi}\lambda^{bcdej}+f^{cbi}\lambda^{abdej}+f^{dbi}\lambda^{acbej}+f^{ebi}\lambda^{acdbj}\,,\\
0&=&f^{abi}\lambda^{bcdjk}+f^{cbi}\lambda^{abdjk}+f^{dbi}\lambda^{acbjk}\,,\\
0&=&f^{abi}\lambda^{bcjkl}+f^{cbi}\lambda^{abjkl}\,,\\
0&=&f^{abi}\lambda^{bjklm}\,,\\
\cr
0&=&f^{abi}\eta^{jbcd}+f^{cbi}\eta^{jabd}+f^{dbi}\eta^{jacb}\,,\\
0&=&f^{abi}\eta^{jkbc}+f^{cbi}\eta^{jkab}\,,\\
0&=&f^{abi}\eta^{jklb}\,,\\
\cr
0&=&f^{abi}\eta^{jbcde}+f^{cbi}\eta^{jabde}+f^{dbi}\eta^{jacbe}+f^{ebi}\eta^{jacdb}\,,\\
0&=&f^{abi}\eta^{jkbcd}+f^{cbi}\eta^{jkabd}+f^{dbi}\eta^{jkacb}\,,\\
0&=&f^{abi}\eta^{jklbc}+f^{cbi}\eta^{jklab}\,,\\
0&=&f^{abi}\eta^{jklmb}\,,\label{jacobi_eta}
\end{eqnarray}
and so on\footnote{Remember here that $\{a,b,c,d,e,f\}$ are off-diagonal indices, while $\{i,j,k,l,m\}$ are diagonal ones.}.\\\\Thus, the gauge-fixing action \eqref{SYM+MAG} is replaced by
\begin{eqnarray}
S_{\mathrm{GSMAG}}&=&\frac{1}{8}s\int dV\,\bigg[c^{a}_{\star}\,\mathcal{D}^{2}\omega^{a}(V)
+c^{i}_{\star}\,\mathcal{D}^{2}\omega^{i}(V)
+\bar{c}^{a}_{\star}\,\bar{\mathcal{D}}^{2}\bar{\omega}^{a}(V)
+\bar{c}^{i}_{\star}\,\bar{\mathcal{D}}^{2}\bar{\omega}^{i}(V)\,
\bigg]\nonumber\\
&=&\frac{1}{8}\int dV\,\bigg\{B^{a}\,\mathcal{D}^{2}\omega^{a}(V)
+B^{i}\,\mathcal{D}^{2}\omega^{i}(V)
+\bar{B}^{a}\,\bar{\mathcal{D}}^{2}\bar{\omega}^{a}(V)
+\bar{B}^{i}\,\bar{\mathcal{D}}^{2}\bar{\omega}^{i}(V)\,
\nonumber\\
&&
-c^{a}_{\star}\,\mathcal{D}^{2}\left[s\omega^{a}(V)\right]
-c^{i}_{\star}\,\mathcal{D}^{2}\left[s\omega^{i}(V)\right]
-\bar{c}^{a}_{\star}\,\bar{\mathcal{D}}^{2}\left[s\bar{\omega}^{a}(V)\right]
-\bar{c}^{i}_{\star}\,\bar{\mathcal{D}}^{2}\left[s\bar{\omega}^{i}(V)\right]
\!\bigg\}\,,
\label{GSMAG_action}
\end{eqnarray}
where $\bar{\omega}^{a,i}(V)$ is the complex conjugate of $\omega^{a,i}(V)$. Naturally, the external sources term must be replaced by
\begin{eqnarray}
\Sigma_{\mathrm{ext}}[\Omega,L,\bar{L},R,P,\bar{R},\bar{P}]&=&s\,\bigg[\,\int dV \,\Big(-\Omega^{a}V^{a}-\Omega^{i}V^{i}\Big)
+\int d^{4}x d^{2}\theta\,\Big(L^{a}c^{a}+L^{i}c^{i}\Big)
\nonumber\\
&&+\int d^{4}x d^{2}\bar\theta\,\Big(\bar{L}^{a}\bar{c}^{a}+\bar{L}^{i}\bar{c}^{i}\Big)
+\int dV\, \Big( R^{a}\omega^{a}(V)+R^{i}\omega^{i}(V)\Big)\nonumber\\
&&+\int dV\, \Big( \bar{R}^{a}\bar{\omega}^{a}(V)+\bar{R}^{i}\bar{\omega}^{i}(V)\Big)\,\bigg]\nonumber\\
&=&\int dV \,\Big[\Omega^{a}\left(sV^{a}\right)+\Omega^{i}\left(sV^{i}\right)\Big]
+\int d^{4}x d^{2}\theta\,\Big[L^{a}\left(sc^{a}\right)+L^{i}\left(sc^{i}\right)\Big]
\nonumber\\
&&+\int d^{4}x d^{2}\bar\theta\,\Big[\bar{L}^{a}\left(s\bar{c}^{a}\right)+\bar{L}^{i}\left(s\bar{c}^{i}\right)\Big]
+\int dV\,\Big\{ P^{a}\omega^{a}(V)\nonumber\\
&&-R^{a}\,\left[s\omega^{a}(V)\right]+P^{i}\omega^{i}(V)-R^{i}\,\left[s\omega^{i}(V)\right]+\bar{P}^{a}\bar{\omega}^{a}(V)\nonumber\\
&&-\bar{R}^{a}\,\left[s\bar{\omega}^{a}(V)\right]+\bar{P}^{i}\bar{\omega}^{i}(V)-\bar{R}^{i}\,\left[s\bar{\omega}^{i}(V)\right]\Big\}\,, 
\label{external_GSMAG}
\end{eqnarray}     
where use has been made of a new set of BRST doublets of external sources
\begin{eqnarray}
&sR^{a,i}=P^{a,i}\,,\qquad sP^{a,i}=0\,,&\nonumber\\
&s\bar{R}^{a,i}=\bar{P}^{a,i}\,,\qquad s\bar{P}^{a,i}=0\,,&
\label{RP_sources}
\end{eqnarray}
with $\{\bar{R},\bar{P}\}$ being the complex conjugated of $\{R,P\}$, respectively. The quantum numbers and the fermionic/bosonic nature of the sources \eqref{RP_sources} are those of the sources $\{R,P\}$ in Table \ref{quantum_numbers}, even for the complex conjugated. Finally, we are in position to replace the action \eqref{S} for a more general one:
\begin{equation}
\Sigma=S_{\mathrm{SYM}}+S_{\mathrm{GSMAG}}+\Sigma_{\mathrm{ext}}[\Omega,L,\bar{L},R,P,\bar{R},\bar{P}]\,.
\label{Sigma}
\end{equation}
The expression \eqref{Sigma} above will be our starting point action, which we will study the renormalizability. This action encodes our previous discussion about the introductory definition of the maximal Abelian gauge in superspace, originally presented in \cite{Capri:2015usa}, and the necessity of a generalization in order to circumvent the ambiguity generated by the dimensionless of the vector superfield. Also, the necessary local composite operators were appropriately defined in \eqref{Sigma}. As we shall see next, a full set of Ward identities can be established for action \eqref{Sigma}. In fact such identities are very similar to the identities (\ref{ST}--\ref{gh_eq}) with few modifications.  

\subsection{The Ward identities for the generalized action}
We display here a full set of Ward identities enjoyed by the action \eqref{Sigma}.
\subsubsection{The (new) Slavnov-Taylor identity}
\begin{eqnarray}
\mathcal{B}(\Sigma)&:=&\int dV\,\left(
\frac{\delta \Sigma}{\delta\Omega^{a}}\frac{\delta \Sigma}{\delta V^{a}}
+\frac{\delta \Sigma}{\delta\Omega^{i}}\frac{\delta \Sigma}{\delta V^{i}}
+P^{a}\frac{\delta \Sigma}{\delta R^{a}}
+P^{i}\frac{\delta \Sigma}{\delta R^{i}}
+\bar{P}^{a}\frac{\delta \Sigma}{\delta \bar{R}^{a}}
+\bar{P}^{i}\frac{\delta \Sigma}{\delta \bar{R}^{i}}\right)\nonumber\\
&&+\int d^{4}x d^{2}\theta\,\left(\frac{\delta \Sigma}{\delta L^{a}}\frac{\delta \Sigma}{\delta c^{a}}
+\frac{\delta \Sigma}{\delta L^{i}}\frac{\delta \Sigma}{\delta c^{i}}
+B^{a}\frac{\delta \Sigma}{\delta c^{a}_{\star}}
+B^{i}\frac{\delta \Sigma}{\delta c^{i}_{\star}}\right)\nonumber\\
&&
+\int d^{4}xd^{2}\bar\theta\,\left(\frac{\delta \Sigma}{\delta \bar{L}^{a}}\frac{\delta \Sigma}{\delta \bar{c}^{a}}
+\frac{\delta\Sigma}{\delta \bar{L}^{i}}\frac{\delta \Sigma}{\delta \bar{c}^{i}}
+\bar{B}^{a}\frac{\delta \Sigma}{\delta \bar{c}^{a}_{\star}}
+\bar{B}^{i}\frac{\delta \Sigma}{\delta \bar{c}^{i}_{\star}}\right)\nonumber\\
&=&0\,.
\label{new_ST}
\end{eqnarray}
\subsubsection{The (new) diagonal gauge-fixing equations}
\begin{eqnarray}
\frac{\delta \Sigma}{\delta B^{i}}-\frac{1}{8}\,\bar{\mathcal{D}}^{2}\mathcal{D}^{2}\frac{\delta\Sigma}{\delta P^{i}}&=&0\,,\label{new_diag_gaugefixing}\\
\cr
\frac{\delta \Sigma}{\delta \bar{B}^{i}}-\frac{1}{8}\,\mathcal{D}^{2}\bar{\mathcal{D}}^{2}\frac{\delta\Sigma}{\delta \bar{P}^{i}}&=&0\,.
\end{eqnarray}
\subsubsection{The (new) off-diagonal gauge-fixing equations}
\begin{eqnarray}
\frac{\delta \Sigma}{\delta B^{a}}-\frac{1}{8}\,{\bar{\mathcal{D}}}^{2}\mathcal{D}^{2}\frac{\delta \Sigma}{\delta P^{a}}&=&0\,,\\
\nonumber\\
\frac{\delta \Sigma}{\delta \bar{B}^{a}}-\frac{1}{8}\,\mathcal{D}^{2}{\bar{\mathcal{D}}}^{2}\frac{\delta \Sigma}{\delta \bar{P}^{a}}&=&0\,.
\label{new_off_diagonal_gaugefixing}
\end{eqnarray}
\subsubsection{The (new) diagonal antighost equations}
\begin{eqnarray}
\frac{\delta\Sigma}{\delta c^{i}_{\star}}-\frac{1}{8}\bar{\mathcal{D}}^{2}\mathcal{D}^{2}\frac{\delta\Sigma}{\delta R^{i}} & = & 0\,,\\\cr
\frac{\delta\Sigma}{\delta\bar{c}^{i}_{\star}}-\frac{1}{8}\mathcal{D}^{2}\bar{\mathcal{D}}^{2}\frac{\delta\Sigma}{\delta \bar{R}^{i}} & = & 0\,.
\end{eqnarray}
\subsubsection{The (new) off-diagonal antighost equations}
\begin{eqnarray}
\frac{\delta\Sigma}{\delta c^{a}_{\star}}-\frac{1}{8}\bar{\mathcal{D}}^{2}\mathcal{D}^{2}\frac{\delta\Sigma}{\delta R^{a}}& = & 0\,,\\\cr
\frac{\delta\Sigma}{\delta\bar{c}^{a}_{\star}}-\frac{1}{8}\mathcal{D}^{2}\bar{\mathcal{D}}^{2}\frac{\delta\Sigma}{\delta \bar{R}^{a}} & = & 0\,.
\end{eqnarray}
\subsubsection{The (new) $\mathcal{R}$-invariance}
\begin{eqnarray}
\mathcal{R}(\Sigma)&:=&
\sum_{X\in\{V,\Omega,R,P,\bar{R},\bar{P}\}}\int dV\,(\delta_{\mathcal{R}} X)\frac{\delta \Sigma}{\delta X}\nonumber\\
&&
+\sum_{Y\in\{c,c_\star,B,L\}}\int d^{4}x d^{2}\theta\,
(\delta_{\mathcal{R}} Y)\frac{\delta \Sigma}{\delta Y}\,\,\,\,\,
+\sum_{\bar{Y}\in\{\bar{c},\bar{c}_{\star},\bar{B},\bar{L}\}}\int d^{4}x d^{2}\bar\theta\,
(\delta_{\mathcal{R}} \bar{Y})\frac{\delta \Sigma}{\delta \bar{Y}}\nonumber\\
&=&0\,,\label{new_R_invariance}
\end{eqnarray}
where the $\mathcal{R}$-variations are given by Eq.~\eqref{R_var}.
\subsubsection{The (new) diagonal rigid invariance}
\begin{eqnarray}
\mathcal{W}^{i}(\Sigma)&:=&
\sum_{X\in\{V,\Omega,R,P,\bar{R},\bar{P}\}}\int dV\,f^{abi}X^{a}\frac{\delta \Sigma}{\delta X^{b}}\nonumber\\
&&
+\sum_{Y\in\{c,c_\star,B,L\}}\int d^{4}x d^{2}\theta\,
f^{abi}Y^{a}\frac{\delta \Sigma}{\delta Y^{b}}\,\,\,\,\,
+\sum_{\bar{Y}\in\{\bar{c},\bar{c}_{\star},\bar{B},\bar{L}\}}\int d^{4}x d^{2}\bar\theta\,
f^{abi}\bar{Y}^{a}\frac{\delta \Sigma}{\delta \bar{Y}^{b}}\nonumber\\
&=&0\,.
\label{new_rigid_symm}
\end{eqnarray}
\subsubsection{The (new) diagonal ghost equation}
\begin{eqnarray}
\mathcal{G}(\Sigma) &:=&\int d^{4}xd^{2}\theta\,\left(\frac{\delta \Sigma}{\delta c^{i}}+f^{abi}c^{a}_{\star}\frac{\delta S}{\delta B^{b}}\right) + \int d^{4}xd^{2}\bar{\theta}\,\left(\frac{\delta \Sigma}{\delta \bar{c}^{i}}+f^{abi}\bar{c}^{a}_{\star}\frac{\delta S}{\delta \bar{B}^{b}}\right)\nonumber\\
&&+\int dV\,f^{abi}\left(R^{a}\frac{\delta \Sigma}{\delta P^{b}}
+\bar{R}^{a}\frac{\delta \Sigma}{\delta \bar{P}^{b}}\right)\nonumber\\
&=&\int d^{4}xd^{2}\theta\,f^{abi}L^{a}c^{b}
+\int d^{4}xd^{2}\bar{\theta}\,f^{abi}\bar{L}^{a}\bar{c}^{b}-\int dV\, f^{abi}\Omega^{a}V^{b}\,.
\label{new_gh_eq}
\end{eqnarray}
The generalized Jacobi identities, Eqs (\ref{jacobi_lambda}--\ref{jacobi_eta}) and generalizations for arbitrary rank tensors, are fundamental in order to establish the ghost equation as written above.\\\\Furthermore, a deeper look at the diagonal and off-diagonal gauge fixing equations, Eqs (\ref{new_diag_gaugefixing}--\ref{new_off_diagonal_gaugefixing}), reveals  that they are very similar and could be written in a more compact way as
\begin{eqnarray}
\frac{\delta \Sigma}{\delta B^{A}}-\frac{1}{8}\,\bar{\mathcal{D}}^{2}\mathcal{D}^{2}\frac{\delta\Sigma}{\delta P^{A}}&=&0\,,\label{new_diag_gaugefixing}\\
\cr
\frac{\delta \Sigma}{\delta \bar{B}^{A}}-\frac{1}{8}\,\mathcal{D}^{2}\bar{\mathcal{D}}^{2}\frac{\delta\Sigma}{\delta \bar{P}^{A}}&=&0\,,
\end{eqnarray}
where $A\equiv\{a,i\}$, i.e. with no difference between the diagonal and off-diagonal sectors. It is possible because we did not give an explicit form for the field functionals \eqref{omega_off} and \eqref{omega_diag}. In fact, the split of these two sectors of the theory is provided here by the rigid symmetry \eqref{new_rigid_symm}, which is exclusively diagonal. An analogous approach was made in  \cite{Capri:2018gpu} in the context of the full Landau gauge. However, in that case, the rigid symmetry extends to the whole SU($n$) group.\\\\As a final comment, we would like to emphasize that the identities displayed above are, in principle, valid only at classical level. At quantum level it is first necessary to prove the absence of anomalies. This is in fact one of the main steps of the algebraic proof of the renormalization. The study of anomalies for such identities in superspace was exhaustively discussed in the literature, see e.g. \cite{Piguet:1981hh,Piguet:1996ys,Piguet:1982qb}, in different gauges. In particular, as we are dealing with a pure super-Yang-Mills theory, the absence of chiral matter fields automatically guarantees the validity of the Slavnoy-Taylor identity \eqref{new_ST} and the $\mathcal{R}$-invariance \eqref{new_R_invariance} at quantum level. Then, we assume from now on that the Ward identities presented here in this section are anomaly free.

\section{Renormalization}
\label{renorm}
Our next step will be to determine the most general invariant counterterm which can be freely added to all order in perturbation theory, allowing us to remove all divergences of the theory. Such counterterm is generically written as
\begin{equation}
\Sigma_{\mathrm{CT}}=\int dV\,\Delta^{(2,0,0)}(x,\theta,\bar{\theta})+\int d^{4}xd^{2}\theta\,\Delta^{(3,0,2)}(x,\theta)
+\int d^{4}xd^{2}\bar{\theta}\,{\Delta}^{(3,0,-2)}(x,\bar{\theta})\,,
\label{counterterm}
\end{equation} 
where $\Delta^{(d,\#g,n)}$ are local polynomials in the fields and sources. The upper labels indicate the mass dimension $(d)$, the ghost number $(\#g)$ and the $\mathcal{R}$-weights $(n)$, respectively, in accordance with Table \ref{quantum_numbers}. Also, the hermiticity condition imposes that the {antichiral} polynomial $\Delta^{(3,0,-2)}$ be the chiral conjugated of the chiral polynomial $\Delta^{(3,0,2)}$.\\\\Then, in order to find an explicit expression for the counterterm, $\Sigma_{\mathrm{CT}}$, we follow the setup of the algebraic renormalization \cite{Piguet:1995er} and perturb the classical action $\Sigma$, Eq.~\eqref{Sigma}, by adding the counterterm described above, demanding that the perturbed
action, $(\Sigma+\epsilon\,\Sigma_{\mathrm{CT}})$, with $\epsilon$ being an expansion parameter, fulfills, to the first order in $\epsilon$, the same Ward identities obeyed by the classical action $\Sigma$, Eqs (\ref{new_ST}--\ref{new_gh_eq}). This amounts to impose the following constraints on $\Sigma_{\mathrm{CT}}$:
\begin{eqnarray}
\mathcal{B}_{\Sigma}(\Sigma_{\mathrm{CT}})&=&0\,,\label{1st_constraint}\\\cr
\frac{\delta \Sigma_{\mathrm{CT}}}{\delta B^{i}}-\frac{1}{8}\,{\bar{\mathcal{D}}}^{2}\mathcal{D}^{2}\frac{\delta \Sigma_{\mathrm{CT}}}{\delta P^{i}}&=&0\,,\label{2nd_constraint}\\\cr
\frac{\delta \Sigma_{\mathrm{CT}}}{\delta \bar{B}^{i}}-\frac{1}{8}\,\mathcal{D}^{2}{\bar{\mathcal{D}}}^{2}\frac{\delta \Sigma_{\mathrm{CT}}}{\delta \bar{P}^{i}}&=&0\,,\\\cr
\frac{\delta \Sigma_{\mathrm{CT}}}{\delta B^{a}}-\frac{1}{8}\,{\bar{\mathcal{D}}}^{2}\mathcal{D}^{2}\frac{\delta \Sigma_{\mathrm{CT}}}{\delta P^{a}}&=&0\,,\\\cr
\frac{\delta \Sigma_{\mathrm{CT}}}{\delta \bar{B}^{a}}-\frac{1}{8}\,\mathcal{D}^{2}{\bar{\mathcal{D}}}^{2}\frac{\delta \Sigma_{\mathrm{CT}}}{\delta \bar{P}^{a}}&=&0\,,\\\cr
\frac{\delta\Sigma_{\mathrm{CT}}}{\delta c^{i}_{\star}}-\frac{1}{8}\bar{\mathcal{D}}^{2}\mathcal{D}^{2}\frac{\delta\Sigma_{\mathrm{CT}}}{\delta R^{i}} & = & 0\,,\\\cr
\frac{\delta\Sigma_{\mathrm{CT}}}{\delta\bar{c}^{i}_{\star}}-\frac{1}{8}\mathcal{D}^{2}\bar{\mathcal{D}}^{2}\frac{\delta\Sigma_{\mathrm{CT}}}{\delta \bar{R}^{i}}& = & 0\,,\\\cr
\frac{\delta\Sigma_{\mathrm{CT}}}{\delta c^{a}_{\star}}-\frac{1}{8}\bar{\mathcal{D}}^{2}\mathcal{D}^{2}\frac{\delta\Sigma_{\mathrm{CT}}}{\delta R^{a}} & = & 0\,,\\\cr
\frac{\delta\Sigma_{\mathrm{CT}}}{\delta\bar{c}^{a}_{\star}}-\frac{1}{8}\mathcal{D}^{2}\bar{\mathcal{D}}^{2}\frac{\delta\Sigma_{\mathrm{CT}}}{\delta \bar{R}^{a}} & = & 0\,,\\\cr
\mathcal{R}(\Sigma_{\mathrm{CT}})&=&0\,,\\\cr
\mathcal{W}^{i}(\Sigma_{\mathrm{CT}})&=&0\,,\\\cr
\mathcal{G}^{i}(\Sigma_{\mathrm{CT}})&=&0\,,\label{last_constraint}
\end{eqnarray}
where $\mathcal{B}_{\Sigma}$ is the so-called nilpotent linearized Slavnov-Taylor operator:
\begin{eqnarray}
\mathcal{B}_{\Sigma} & = & \int dV\,\Bigg(\frac{\delta\Sigma}{\delta\Omega^{a}}\frac{\delta}{\delta V^{a}}+\frac{\delta\Sigma}{\delta V^{a}}\frac{\delta}{\delta\Omega^{a}}+\frac{\delta\Sigma}{\delta\Omega^{i}}\frac{\delta}{\delta V^{i}}+\frac{\delta\Sigma}{\delta V^{i}}\frac{\delta}{\delta\Omega^{i}}\Bigg)\nonumber\\
&&+\int dV\left(P^{a}\frac{\delta}{\delta R^{a}}+P^{i}\frac{\delta}{\delta R^{i}}
+\bar{P}^{a}\frac{\delta}{\delta \bar{R}^{a}}+\bar{P}^{i}\frac{\delta}{\delta \bar{R}^{i}}\right)\nonumber\\
&&+  \int d^{4}xd^{2}\theta\,\Bigg(\frac{\delta\Sigma}{\delta L^{a}}\frac{\delta}{\delta c^{a}}+\frac{\delta\Sigma}{\delta c^{a}}\frac{\delta}{\delta L^{a}}+\frac{\delta\Sigma}{\delta L^{i}}\frac{\delta}{\delta c^{i}}+\frac{\delta\Sigma}{\delta c^{i}}\frac{\delta}{\delta L^{i}}+B^{a}\frac{\delta}{\delta c_{\star}^{a}}+B^{i}\frac{\delta}{\delta c_{\star}^{i}}\Bigg)\nonumber\\
&&+ \int d^{4}xd^{2}\bar{\theta}\,\left(\frac{\delta\Sigma}{\delta\bar{L}^{a}}\frac{\delta}{\delta\bar{c}^{a}}+\frac{\delta\Sigma}{\delta\bar{c}^{a}}\frac{\delta}{\delta\bar{L}^{a}}+\frac{\delta\Sigma}{\delta\bar{L}^{i}}\frac{\delta}{\delta\bar{c}^{i}}+\frac{\delta\Sigma}{\delta\bar{c}^{i}}\frac{\delta}{\delta\bar{L}^{i}}+\bar{B}^{a}\frac{\delta}{\delta\bar{c}_{\star}^{a}}+\bar{B}^{i}\frac{\delta}{\delta\bar{c}_{\star}^{i}}\right)\,.
\label{ST_linearized}
\end{eqnarray}
The presence of the nilpotent operator $\mathcal{B}_{\Sigma}$ transforms the problem of obtain the counterterm in a cohomology problem of the operator $\mathcal{B}_{\Sigma}$. In fact, $\Sigma_{\mathrm{CT}}$ can be written as
\begin{equation}
\Sigma_{\mathrm{CT}}=a_0\,S_{\mathrm{SYM}}+\mathcal{B}_{\Sigma}\Delta^{(0,-1,0)}\,,
\label{CT}
\end{equation}
with $a_0$ an arbitrary real coefficient and with $\Delta^{(0,-1,0)}$ given by
\begin{equation}
\Delta^{(0,-1,0)}=\int dV\,\Delta^{(2,-1,0)}(x,\theta,\bar{\theta})+\int d^{4}xd^{2}\theta\,\Delta^{(3,-1,2)}(x,\theta)+\int d^{4}xd^{2}\bar{\theta}\,\Delta^{(3,-1,-2)}(x,\bar{\theta})\,,
\end{equation}
since $S_{\mathrm{SYM}}$ cannot be written as an exact $\mathcal{B}_{\Sigma}$ variation. The remaining constrains, Eqs~(\ref{2nd_constraint}--\ref{last_constraint}), provide that $\Delta^{(0,-1,0)}$ is written as
\begin{eqnarray}
\Delta^{(0,-1,0)} & = & \int dV\,\bigg[\,F^{i}(V)\,\Omega^{i}+F^{a}(V)\,\Omega^{a}
+\,G^{i}(V)\,\bigg(R^{i}+\frac{1}{8}\,\mathcal{D}^{2}c_{\star}^{i}\bigg)+G^{a}(V)\,\bigg(R^{a}+\frac{1}{8}\,\mathcal{D}^{2}c_{\star}^{a}\bigg)\nonumber \\
&&+\,\bar{G}^{i}(V)\,\bigg(\bar{R}^{i}+\frac{1}{8}\,\bar{\mathcal{D}}^{2}\bar{c}_{\star}^{i}\bigg)+\bar{G}^{a}(V)\,\bigg(\bar{R}^{a}+\frac{1}{8}\,\bar{\mathcal{D}}^{2}\bar{c}_{\star}^{a}\bigg)\bigg]\nonumber\\
 && +  \int d^{4}xd^{2}\theta\,\,a_{1}L^{a}c^{a}
 +  \int d^{4}xd^{2}\bar{\theta}\,\,\bar{a}_{1}\bar{L}^{a}\bar{c}^{a}\,,
 \label{Delta_minus_one}
\end{eqnarray}
where, $\{a_1,\bar{a}_{1}\}$ is a pair of complex conjugated arbitrary coefficients and thanks to the dimensionless of the vector superfield, $F^{a,i}(V)$ and $G^{a,i}(V)$ are power series in $V$:
\begin{eqnarray}
F^{a}(V) & = & \alpha_1\,V^{a}+\alpha_2\, f^{abi}V^{b}V^{i}+\sum_{n=3}^{\infty}\alpha^{aA_1A_2\dots A_{n}}\,V^{A_1}V^{A_2}\dots V^{A_{n}}\,,\label{Fa}\\
\cr
F^{i}(V)& = & \beta\,V^{i}+\sum_{n=3}^{\infty}\beta^{iA_1A_2\dots A_{n}}\,V^{A_1}V^{A_2}\dots V^{A_{n}}\,,\label{Fi}\\\cr
G^{a}(V) & = & \kappa_1\,V^{a}+\kappa_2\, f^{abi}V^{b}V^{i}+\sum_{n=3}^{\infty}\kappa^{aA_1A_2\dots A_{n}}\,V^{A_1}V^{A_2}\dots V^{A_{n}}\,,\label{Ga}\\
\cr
G^{i}(V)& = & \sigma\,V^{i}+\sum_{n=3}^{\infty}\sigma^{iA_1A_2\dots A_{n}}\,V^{A_1}V^{A_2}\dots V^{A_{n}}\,.
\label{Gi}
\end{eqnarray}
Naturally, $\bar{G}^{a,i}(V)$ are the complex conjugated of $G^{a,i}(V)$ and the rank-$n$ tensors: $\alpha^{aA_1A_2\dots A_{n}}$, $\beta^{iA_1A_2\dots A_{n}}$, $\kappa^{aA_1A_2\dots A_{n}}$ and $\sigma^{iA_1A_2\dots A_{n}}$ must obey generalized Jacobi identities similar to (\ref{jacobi_lambda}--\ref{jacobi_eta}).\\\\Acting the linearized operator $\mathcal{B}_{\Sigma}$ on $\Delta^{(0,-1,0)}$ in \eqref{Delta_minus_one}, it is possible to obtain  explicit expressions for the local polynomials $\Delta^{(d,\#g,n)}$ in \eqref{counterterm}:
\begin{eqnarray}
\Delta^{(2,0,0)}&\!\!\!=\!\!\!&\left(\frac{\delta\Sigma}{\delta\Omega^{j}}\frac{\partial F^{i}(V)}{\partial V^{j}}
+\frac{\delta\Sigma}{\delta\Omega^{a}}\frac{\partial F^{i}(V)}{\partial V^{a}}\right)\Omega^{i}
+F^{i}(V)\frac{\delta\Sigma}{\delta V^{i}}\cr\cr
&&+\left(\frac{\delta\Sigma}{\delta\Omega^{i}}\frac{\partial F^{a}(V)}{\partial V^{i}}
+\frac{\delta\Sigma}{\delta\Omega^{b}}\frac{\partial F^{a}(V)}{\partial V^{b}}\right)\Omega^{a}
+F^{a}(V)\frac{\delta\Sigma}{\delta V^{a}}\cr\cr
&&+\left(\frac{\delta\Sigma}{\delta\Omega^{j}}\frac{\partial G^{i}(V)}{\partial V^{j}}
+\frac{\delta\Sigma}{\delta\Omega^{a}}\frac{\partial G^{i}(V)}{\partial V^{a}}\right)\! \left(R^{i}+\frac{1}{8}\mathcal{D}^{2}c^{i}_{\star}\right)
+G^{i}(V)P^{i}+\frac{1}{8}\,B^{i}\mathcal{D}^{2}G^{i}(V)\cr\cr
&&+\left(\frac{\delta\Sigma}{\delta\Omega^{i}}\frac{\partial G^{a}(V)}{\partial V^{i}}
+\frac{\delta\Sigma}{\delta\Omega^{b}}\frac{\partial G^{a}(V)}{\partial V^{b}}\right)\! \left(R^{a}+\frac{1}{8}\mathcal{D}^{2}c^{a}_{\star}\right)
+G^{a}(V)P^{a}+\frac{1}{8}\,B^{a}\mathcal{D}^{2}G^{a}(V)\cr\cr
&&+\left(\frac{\delta\Sigma}{\delta\Omega^{j}}\frac{\partial \bar{G}^{i}(V)}{\partial V^{j}}
+\frac{\delta\Sigma}{\delta\Omega^{a}}\frac{\partial \bar{G}^{i}(V)}{\partial V^{a}}\right)\! \left(\bar{R}^{i}+\frac{1}{8}\bar{\mathcal{D}}^{2}\bar{c}^{i}_{\star}\right)
+\bar{G}^{i}(V)\bar{P}^{i}+\frac{1}{8}\,\bar{B}^{i}\bar{\mathcal{D}}^{2}\bar{G}^{i}(V)\cr\cr
&&+\left(\frac{\delta\Sigma}{\delta\Omega^{i}}\frac{\partial \bar{G}^{a}(V)}{\partial V^{i}}
+\frac{\delta\Sigma}{\delta\Omega^{b}}\frac{\partial \bar{G}^{a}(V)}{\partial V^{b}}\right)\! \left(\bar{R}^{a}+\frac{1}{8}\bar{\mathcal{D}}^{2}\bar{c}^{a}_{\star}\right)
+\bar{G}^{a}(V)\bar{P}^{a}+\frac{1}{8}\,\bar{B}^{a}\bar{\mathcal{D}}^{2}\bar{G}^{a}(V)\,,\nonumber\\\\\cr
\Delta^{(3,0,2)}&=&-\frac{a_{0}}{128g^{2}}(W^{a\alpha}W^{a}_{\alpha}+{W}^{i\alpha}{W}^{i}_{\alpha})
+a_{1}\,\frac{\delta\Sigma}{\delta c^{a}}c^{a}+a_{1}\,L^{a}\frac{\delta\Sigma}{\delta L^{a}}\,,\\\cr
\Delta^{(3,0,-2)}&=&-\frac{a_{0}}{128g^{2}}(\bar{W}^{a}_{\dot{\alpha}}\bar{W}^{a\dot{\alpha}}+\bar{W}^{i}_{\dot{\alpha}}\bar{W}^{i\dot{\alpha}})
+\bar{a}_{1}\,\frac{\delta\Sigma}{\delta \bar{c}^{a}}\bar{c}^{a}+\bar{a}_{1}\,\bar{L}^{a}\frac{\delta\Sigma}{\delta \bar{L}^{a}}\,.
\end{eqnarray}
Furthermore, it is useful to write the counterterm in a parametric form. Then, in order to establish such expression let us first notice that the nontrivial part of the countertem can be easily rewritten as
\begin{eqnarray}
a_0\,S_{\mathrm{SYM}} & = & -\frac{a_{0}}{128g^{2}}\int d^{4}xd^{2}\theta\,W^{A\alpha}W^{A}_{\alpha}+ \mathrm{c.c.}=-a_{0}\,g^{2}\frac{\partial S_{\mathrm{SYM}}}{\partial g^{2}}=-a_{0}\,g^{2}\frac{\partial\Sigma}{\partial g^{2}}\,,
\label{parametric_SYM}
\end{eqnarray}
while the remaining terms of \eqref{CT}, i.e. the so-called trivial part of the cohomology is written in the same fashion. Namely,
\begin{eqnarray}
\mathcal{B}_{\Sigma}\Delta^{(0,-1,0)} &\!\!\! =\!\!\! & \int dV\,\left(F^{a}(V)\frac{\delta\Sigma}{\delta V^{a}}+\kappa_{1}R^{a}\frac{\delta\Sigma}{\delta R^{a}}+\kappa_{1}P^{a}\frac{\delta\Sigma}{\delta P^{a}}
+\bar{\kappa}_{1}\bar{R}^{a}\frac{\delta\Sigma}{\delta \bar{R}^{a}}+\bar{\kappa}_{1}\bar{P}^{a}\frac{\delta\Sigma}{\delta \bar{P}^{a}}\right)\cr\cr
&& -\int dV\,\left(\Omega^{b}\frac{\partial F^{b}(V)}{\partial V^{a}}+\Omega^{i}\frac{\partial F^{i}(V)}{\partial V^{a}}
\right)\frac{\delta\Sigma}{\delta\Omega^{a}}\cr\cr
&& + \int dV\,\left(F^{i}(V)\frac{\delta\Sigma}{\delta V^{i}}+\sigma\,R^{i}\frac{\delta\Sigma}{\delta R^{i}}+\sigma\,P^{i}\frac{\delta\Sigma}{\delta P^{i}}
 +\bar{\sigma}\,\bar{R}^{i}\frac{\delta\Sigma}{\delta \bar{R}^{i}}+\bar{\sigma}\,\bar{P}^{i}\frac{\delta\Sigma}{\delta \bar{P}^{i}}\right)\cr\cr
&& - \int dV\,\left(\Omega^{a}\frac{\partial F^{a}(V)}{\partial V^{i}}+\Omega^{j}\frac{\partial F^{j}(V)}{\partial V^{i}}\right)\frac{\delta\Sigma}{\delta\Omega^{i}}\cr\cr
&& + \int d^{4}xd^{2}\theta\,\left( a_{1}\,L^{a}\frac{\delta\Sigma}{\delta L^{a}}-a_{1}\,c^{a}\frac{\delta\Sigma}{\delta c^{a}}
+\kappa_{1}\,c^{a}_{\star}\frac{\delta\Sigma}{\delta c_{\star}^{a}}
+\kappa_{1}\,B^{a}\frac{\delta\Sigma}{\delta B^{a}}+\sigma\,c^{i}_{\star}\frac{\delta\Sigma}{\delta c_{\star}^{i}}+\sigma\,B^{i}\frac{\delta\Sigma}{\delta B^{i}}\right)\cr\cr
&& +\int d^{4}xd^{2}\bar{\theta}\,\left( \bar{a}_{1}\,\bar{L}^{a}\frac{\delta\Sigma}{\delta\bar{L}^{a}}-\bar{a}_{1}\,\bar{c}^{a}\frac{\delta\Sigma}{\delta\bar{c}^{a}}
+\bar{\kappa}_{1}\,\bar{c}_{\star}^{a}\frac{\delta\Sigma}{\delta\bar{c}_{\star}^{a}}
+\bar{\kappa}_{1}\,\bar{B}^{a}\frac{\delta\Sigma}{\delta\bar{B}^{a}}+\bar{\sigma}\,\bar{c}_{\star}^{i}\frac{\delta\Sigma}{\delta\bar{c}_{\star}^{i}}+\bar{\sigma}\,\bar{B}^{i}\frac{\delta\Sigma}{\delta\bar{B}^{i}}\right) \cr\cr
&& + (\kappa_{2}-\kappa_{1}\lambda)\frac{\partial\Sigma}{\partial \lambda}
 +\sum_{n=3}^{\infty}\,\left(\kappa^{aA_{1}A_{2}\dots A_{n}}-\kappa_{1}\lambda^{aA_{1}A_{2}\dots A_{n}}\right)\frac{\partial\Sigma}{\partial \lambda^{aA_{1}A_{2}\dots A_{n}}}
\cr\cr
&& + (\bar{\kappa}_{2}-\bar{\kappa}_{1}\bar{\lambda})\frac{\partial\Sigma}{\partial \bar{\lambda}}
+\sum_{n=3}^{\infty}\,\left(\bar{\kappa}^{aA_{1}A_{2}\dots A_{n}}-\bar{\kappa}_{1}\bar{\lambda}^{aA_{1}A_{2}\dots A_{n}}\right)\frac{\partial\Sigma}{\partial \bar{\lambda}^{aA_{1}A_{2}\dots A_{n}}}
\cr\cr
&& + \sum_{n=3}^{\infty}\,\left(\sigma^{iA_{1}A_{2}\dots A_{n}}-\sigma\,\eta^{iA_{1}A_{2}\dots A_{n}}\right)\frac{\partial\Sigma}{\partial \eta^{iA_{1}A_{2}\dots A_{n}}}
\cr\cr
&& + \sum_{n=3}^{\infty}\,\left(\bar{\sigma}^{iA_{1}A_{2}\dots A_{n}}-\bar{\sigma}\,\bar{\eta}^{iA_{1}A_{2}\dots A_{n}}\right)\frac{\partial\Sigma}{\partial \bar{\eta}^{iA_{1}A_{2}\dots A_{n}}}\,.
\label{parametric}
\end{eqnarray}
Combining \eqref{parametric_SYM} and \eqref{parametric} the counterterm can be viewed as 
\begin{eqnarray}
\Sigma_{\mathrm{CT}}=\mathcal{O}\Sigma\,,
\label{CT_equal_to_OSigma}
\end{eqnarray}
where $\mathcal{O}$ is a linear operator acting on $\Sigma$ being given by
\begin{eqnarray}
\mathcal{O}&\!\!\! =\!\!\! &-a_{0}\,g^{2}\frac{\partial}{\partial g^{2}}+\int dV\,\left(F^{a}(V)\frac{\delta}{\delta V^{a}}+\kappa_{1}R^{a}\frac{\delta}{\delta R^{a}}+\kappa_{1}P^{a}\frac{\delta}{\delta P^{a}}
+\bar{\kappa}_{1}\bar{R}^{a}\frac{\delta}{\delta \bar{R}^{a}}+\bar{\kappa}_{1}\bar{P}^{a}\frac{\delta}{\delta \bar{P}^{a}}\right)\cr\cr
&& -\int dV\,\left(\Omega^{b}\frac{\partial F^{b}(V)}{\partial V^{a}}+\Omega^{i}\frac{\partial F^{i}(V)}{\partial V^{a}}
\right)\frac{\delta}{\delta\Omega^{a}}\cr\cr
&& + \int dV\,\left(F^{i}(V)\frac{\delta}{\delta V^{i}}+\sigma\,R^{i}\frac{\delta}{\delta R^{i}}+\sigma\,P^{i}\frac{\delta}{\delta P^{i}}
 +\bar{\sigma}\,\bar{R}^{i}\frac{\delta}{\delta \bar{R}^{i}}+\bar{\sigma}\,\bar{P}^{i}\frac{\delta}{\delta \bar{P}^{i}}\right)\cr\cr
&& - \int dV\,\left(\Omega^{a}\frac{\partial F^{a}(V)}{\partial V^{i}}+\Omega^{j}\frac{\partial F^{j}(V)}{\partial V^{i}}\right)\frac{\delta}{\delta\Omega^{i}}\cr\cr
&& + \int d^{4}xd^{2}\theta\,\left( a_{1}\,L^{a}\frac{\delta}{\delta L^{a}}-a_{1}\,c^{a}\frac{\delta}{\delta c^{a}}
+\kappa_{1}\,c^{a}_{\star}\frac{\delta}{\delta c_{\star}^{a}}
+\kappa_{1}\,B^{a}\frac{\delta}{\delta B^{a}}+\sigma\,c^{i}_{\star}\frac{\delta}{\delta c_{\star}^{i}}+\sigma\,B^{i}\frac{\delta}{\delta B^{i}}\right)\cr\cr
&& +\int d^{4}xd^{2}\bar{\theta}\,\left( \bar{a}_{1}\,\bar{L}^{a}\frac{\delta}{\delta\bar{L}^{a}}-\bar{a}_{1}\,\bar{c}^{a}\frac{\delta}{\delta\bar{c}^{a}}
+\bar{\kappa}_{1}\,\bar{c}_{\star}^{a}\frac{\delta}{\delta\bar{c}_{\star}^{a}}
+\bar{\kappa}_{1}\,\bar{B}^{a}\frac{\delta}{\delta\bar{B}^{a}}+\bar{\sigma}\,\bar{c}_{\star}^{i}\frac{\delta}{\delta\bar{c}_{\star}^{i}}+\bar{\sigma}\,\bar{B}^{i}\frac{\delta}{\delta\bar{B}^{i}}\right) \cr\cr
&& + (\kappa_{2}-\kappa_{1}\lambda)\frac{\partial}{\partial \lambda}
 +\sum_{n=3}^{\infty}\,\left(\kappa^{aA_{1}A_{2}\dots A_{n}}-\kappa_{1}\lambda^{aA_{1}A_{2}\dots A_{n}}\right)\frac{\partial}{\partial \lambda^{aA_{1}A_{2}\dots A_{n}}}
\cr\cr
&& + (\bar{\kappa}_{2}-\bar{\kappa}_{1}\bar{\lambda})\frac{\partial}{\partial \bar{\lambda}}
+\sum_{n=3}^{\infty}\,\left(\bar{\kappa}^{aA_{1}A_{2}\dots A_{n}}-\bar{\kappa}_{1}\bar{\lambda}^{aA_{1}A_{2}\dots A_{n}}\right)\frac{\partial}{\partial \bar{\lambda}^{aA_{1}A_{2}\dots A_{n}}}
\cr\cr
&& + \sum_{n=3}^{\infty}\,\left(\sigma^{iA_{1}A_{2}\dots A_{n}}-\sigma\,\eta^{iA_{1}A_{2}\dots A_{n}}\right)\frac{\partial}{\partial \eta^{iA_{1}A_{2}\dots A_{n}}}
\cr\cr
&& + \sum_{n=3}^{\infty}\,\left(\bar{\sigma}^{iA_{1}A_{2}\dots A_{n}}-\bar{\sigma}\,\bar{\eta}^{iA_{1}A_{2}\dots A_{n}}\right)\frac{\partial}{\partial \bar{\eta}^{iA_{1}A_{2}\dots A_{n}}}\,.
\label{O}
\end{eqnarray}
Now, in order to find the renormalization of the fields, sources and parameters of the theory, let us first define a variable $\Phi$ representing all these quantities, i.e.
\begin{equation}
\Phi\equiv V,B,\bar{B},c,\bar{c},c_{\star},\bar{c}_{\star},\Omega,L,\bar{L},R,\bar{R},P,\bar{P},g,\lambda,\bar{\lambda},\lambda^{aA_{1}A_{2}\dots},\bar{\lambda}^{aA_{1}A_{2}\dots},\eta^{iA_{1}A_{2}\dots},\bar{\eta}^{aA_{1}A_{2}\dots}\,.
\end{equation}
Thus, the bare quantities $\Phi_0$ are related to the renormalized ones by
\begin{equation}
\Phi_0=\Phi+\epsilon\,\zeta_{\Phi}+O(\epsilon^{2})\,,
\end{equation} 
where $\zeta_{\Phi}$ is assumed to have a general form of a local functional of the fields. This assumption is necessary in view of the possibility of nonlinear and/or matricial renormalizations\footnote{In its simplest form, $\zeta_{\Phi}$ is proportional to its corresponding $\Phi$ times a constant.}. Then, the bare classical action $\Sigma[\Phi_0]$ is related to the renormalized action $\Sigma[\Phi]$ as
\begin{equation}
\Sigma[\Phi_{0}]=\Sigma[\Phi+\epsilon\,\zeta_{\Phi}]=\Sigma[\Phi]+\epsilon\,\zeta_{\Phi}\frac{\mathrm{d}\Sigma}{\mathrm{d}\Phi}+O(\epsilon^{2})\,,
\end{equation}
where the derivative $\mathrm{d}/\mathrm{d}\Phi$ is a partial derivative when $\Phi$ represents a parameter and an integrated functional derivative when $\Phi$ represents a field or a source. Now, the expression above can be compared with
\begin{equation}
\Sigma[\Phi_0]=\Sigma[\Phi]+\epsilon\,\Sigma_{\mathrm{CT}}+O(\epsilon^{2})=\Sigma[\Phi]+\epsilon\,\mathcal{O}\Sigma+O(\epsilon^{2})\,,
\end{equation}
which comes from perturbation theory and from \eqref{CT_equal_to_OSigma}. Therefore,
\begin{equation}
\mathcal{O}\equiv \zeta_{\Phi}\frac{\mathrm{d}}{\mathrm{d}\Phi}\,.
\end{equation}
Then, the counterterm can be reabsorbed in the starting point action by the following renormalizations 
\begin{equation}
\Phi_0=\Phi+\epsilon\,\mathcal{O}\Phi+O(\epsilon^{2})\,.
\end{equation}
Taking the last expression into account, the components of the vector superfield renormalize, up to order $\epsilon$, as
\begin{eqnarray}
V^{a}_{0}&=&V^{a}+\epsilon\,F^{a}(V)\,,\\\cr
V^{i}_{0}&=&V^{i}+\epsilon\,F^{i}(V)\,,
\end{eqnarray}
while the sources $\Omega^{a,i}$ renormalize as
\begin{eqnarray}
\Omega^{a}_{0}&=&\Omega^{a}-\epsilon\,\left(\Omega^{b}\frac{\partial F^{b}(V)}{\partial V^{a}}+\Omega^{i}\frac{\partial F^{i}(V)}{\partial V^{a}}\right)\,,
\label{Omega_0_off}\\\cr
\Omega^{i}_{0}&=&\Omega^{i}-\epsilon\,\left(\Omega^{a}\frac{\partial F^{a}(V)}{\partial V^{i}}+\Omega^{j}\frac{\partial F^{j}(V)}{\partial V^{i}}\right)\,.
\label{Omega_0_diag}
\end{eqnarray}
The remaining fields and sources renormalize as 
\begin{eqnarray}
c^{a}_{0}&=&Z\,\,c^{a}\,,\\\cr
L^{a}_{0}&=&Z^{-1}\,\,L^{a}\,,\\\cr
\bar{c}^{a}_{0}&=&\bar{Z}\,\,\bar{c}^{a}\,,\\\cr
\bar{L}^{a}_{0}&=&\bar{Z}^{-1}\,\,\bar{L}^{a}\,,\\\cr
(c^{a}_{\star},B^{a},R^{a},P^{a})_{0}&=&Z_{\star}\,\,(c^{a}_{\star},B^{a},R^{a},P^{a})\,,\\\cr
(\bar{c}^{a}_{\star},\bar{B}^{a},\bar{R}^{a},\bar{P}^{a})_{0}&=&\bar{Z}_{\star}\,\,(\bar{c}^{a}_{\star},\bar{B}^{a},\bar{R}^{a},\bar{P}^{a})\,,\\\cr
(c^{i}_{\star},B^{i},R^{i}, P^{i})_{0}&=&\mathcal{Z}_{\star}\,\,(c^{i}_{\star},B^{i},R^{i}, P^{i})\,,\\\cr
(\bar{c}^{i}_{\star},\bar{B}^{i},\bar{R}^{i},\bar{P}^{i})_{0}&=&\bar{\mathcal{Z}}_{\star}\,\,(\bar{c}^{i}_{\star},\bar{B}^{i},\bar{R}^{i},\bar{P}^{i})\,,\\\cr
(c^{i},\bar{c}^{i},L^{i},\bar{L^{i}})_{0}&=&(c^{i},\bar{c}^{i},L^{i},\bar{L^{i}})\,,
\end{eqnarray}
with
\begin{eqnarray}
Z&=&1-\epsilon\,a_{1}\,,\\\cr
Z^{-1}&=&1+\epsilon\,a_{1}\,,\\\cr
\bar{Z}&=&1-\epsilon\,\bar{a}_{1}\,,\\\cr
\bar{Z}^{-1}&=&1+\epsilon\,\bar{a}_{1}\,,\\\cr
Z_{\star}&=&1+\epsilon\,\kappa_1\,,\\\cr
\bar{Z}_{\star}&=&1+\epsilon\,\bar{\kappa}_1\,,\\\cr
\mathcal{Z}_{\star}&=&1+\epsilon\,\sigma\,,\\\cr
\bar{\mathcal{Z}}_{\star}&=&1+\epsilon\,\bar{\sigma}\,.\\\cr
\end{eqnarray}
Finally, the parameters renormalize as
\begin{eqnarray}
g_0&=&Z_g\,g\,\,=\,\,(1-\epsilon\,a_{0})\,g\,,\\\cr
\lambda_{0}&=&\lambda+\epsilon\,(\kappa_2-\kappa_1 \lambda)\,,\\\cr
\lambda_{0}^{aA_{1}A_{2}\dots}&=&\lambda^{aA_{1}A_{2}\dots}+\epsilon\,(\kappa^{aA_{1}A_{2}\dots}-\kappa_1\lambda^{aA_{1}A_{2}\dots})\,,\\\cr
\bar{\lambda}_{0}^{aA_{1}A_{2}\dots}&=&\bar{\lambda}^{aA_{1}A_{2}\dots}+\epsilon\,(\bar{\kappa}^{aA_{1}A_{2}\dots}-\bar{\kappa}_1\bar{\lambda}^{aA_{1}A_{2}\dots})\,,\\\cr
\eta_{0}^{aA_{1}A_{2}\dots}&=&\eta^{aA_{1}A_{2}\dots}+\epsilon\,(\sigma^{aA_{1}A_{2}\dots}-\sigma\,\eta^{aA_{1}A_{2}\dots})\,,\\\cr
\bar{\eta}_{0}^{aA_{1}A_{2}\dots}&=&\bar{\eta}^{aA_{1}A_{2}\dots}+\epsilon\,(\bar{\sigma}^{aA_{1}A_{2}\dots}-\bar{\sigma}\,\bar{\eta}^{aA_{1}A_{2}\dots})\,.
\end{eqnarray}
These expressions end the proof of the renormalizability of the theory, but some comments about the expressions above are necessary. First, notice that the renormalizations of the components of the vector superfield are nonlinear as $F^{a,i}(V)$ are power series in $V$, as stated in Eqs \eqref{Fa} and \eqref{Fi}. It is also clear from Eqs \eqref{Fa} and \eqref{Fi} that the diagonal and off-diagonal components of $V$ are mixed. In other words, the renormalization between $V^{a}$ and $V^{i}$ is matricial. It can be put in a clear way by noticing that the power series $F^{a,i}(V)$ can always be written as
\begin{eqnarray}
F^{a}(V)&=&\alpha_{1}V^{a}+F^{ab}(V)V^{b}+F^{ai}(V)V^{i}\,,\\\cr
F^{i}(V)&=&\beta\,V^{i}+F^{ia}(V)V^{a}+F^{ij}(V)V^{j}\,,
\end{eqnarray}
with $F^{ai}(V)\neq F^{ia}(V)$ in general. Then we have the following matricial renormalization:
\begin{equation}
\left(\begin{matrix}
V^{a}_{0}\cr\cr
V^{i}_{0}
\end{matrix}\right)
=
\left(\begin{matrix}
Z^{ab}_{V}&Z^{aj}_{V}\cr\cr
Z^{ib}_{V}&Z^{ij}_{V}
\end{matrix}\right)
\left(\begin{matrix}
V^{b}\cr\cr
V^{j}
\end{matrix}\right)\,,
\end{equation}
where
\begin{eqnarray}
Z^{ab}_{V}&=&\delta^{ab}+\epsilon\,[\alpha_{1}\delta^{ab}+F^{ab}(V)]\,,\\\cr
Z^{aj}_{V}&=&\epsilon\,F^{aj}(V)\,,\\\cr
Z^{ib}_{V}&=&\epsilon\,F^{ib}(V)\,,\\\cr
Z^{ij}_{V}&=&\delta^{ij}+\epsilon\,[\beta\,\delta^{ij}+F^{ij}(V)]\,.
\end{eqnarray}
The expressions \eqref{Omega_0_off} and \eqref{Omega_0_diag} for the renormalizations of the sources $\Omega^{a,i}$ also indicate matricial and nonlinear renormalizations:
\begin{equation}
\left(\begin{matrix}
\Omega^{a}_{0}\cr\cr
\Omega^{i}_{0}
\end{matrix}\right)=
\left(\begin{matrix}
Z^{ab}_{\Omega}&Z^{aj}_{\Omega}\cr\cr
Z^{ib}_{\Omega}&Z^{ij}_{\Omega}
\end{matrix}\right)
\left(\begin{matrix}
\Omega^{b}\cr\cr
\Omega^{j}
\end{matrix}\right)\,,
\end{equation}
where
\begin{eqnarray}
Z_{\Omega}^{ab}&=&\delta^{ab}-\epsilon\,\frac{\partial F^{b}(V)}{\partial V^{a}}\,,\\\cr
Z^{aj}_{\Omega}&=&-\epsilon\,\frac{\partial F^{j}(V)}{\partial V^{a}}\,,\\\cr
Z^{ib}_{\Omega}&=&-\epsilon\,\frac{\partial F^{b}(V)}{\partial V^{i}}\,,\\\cr
Z^{ij}_{\Omega}&=&\delta^{ij}-\epsilon\,\frac{\partial F^{j}(V)}{\partial V^{i}}\,.
\end{eqnarray}
After the removal of the divergencies of the theory, the original SSMAG can now be reobtained by choosing values for the gauge parameters. In fact, the particular choices
\begin{equation}
\lambda =-\frac{i}{2}\,,\qquad
\bar{\lambda}=+\frac{i}{2}\,,\qquad
\lambda^{aA_{1}A_{2}\dots}=\bar{\lambda}^{aA_{1}A_{2}\dots}=\eta^{iA_{1}A_{2}\dots}=\bar{\eta}^{iA_{1}A_{2}\dots}=0\,,\\
\end{equation}  
lead us from the GSMAG to the SSMAG.\\\\A final comment emerges from a comparison between the supersymmetric and ordinary cases. In the study of the renormalization of the Yang-Mills action quantized in the maximal Abelian gauge, the absence of the off-diagonal {gauge-fixing} and antighost equations as genuine Ward identities gives rise to extra interaction terms among the ghosts fields. In fact, quartic interaction ghost terms naturaly emerges, as a diagrammatic analisys reveals, and the original gauge can only be defined modulo an extra gauge parameter \cite{Dudal:2004rx}. In the earlier paper \cite{Capri:2015usa}, when the SSMAG was first presented, we also proposed possible quartic interaction ghost terms following the non-supersymmetric approach. However, in that occasion we did not realize that the off-diagonal gauge fixing and antighost equations, given by Eqs \eqref{chiral_off_diag_gaugefixing_eq}, \eqref{antichiral_off_diag_gaugefixing_eq}, \eqref{chiral_off_diag_antighost_eq} and \eqref{antichiral_off_diag_antighost_eq} for the SSMAG, could be established in the supersymmetric scenario, neither that the general approach, given by the GSMAG, should be implemented.

\section{Conclusions and perspectives}
\label{conclusions}
In this work, we have concluded the algebraic proof of the renormazibility of a $N=1$ super-Yang-Mills theory for SU($n$) group in a supersymmetric version of the maximal Abelian gauge. A generalized version of the original propose, Ref.~\cite{Capri:2015usa}, has been adopted. We called this extended version as GSMAG (Generalized-Super-Maximal-Abelian-Gauge). Such version depends on a set of infinity gauge parameters but, at the end, the original version, called SSMAG (Simplest-Super-Maximal-Abelian-Gauge), can be achieved from the generalized one by a suitable adjusting of the gauge parameters, which are, however, fundamental in the algebraic proof.\\\\The proof presented here is very similar to that one presented in \cite{Capri:2018gpu} in the case of the Landau gauge. The main difference is that the gauge symmetry group is explicitly split into its diagonal and off-diagonal parts. This split is made evident from the diagonal rigid symmetry \eqref{new_rigid_symm} and the consequent generalized Jacobi identities (\ref{jacobi_lambda}--\ref{jacobi_eta}) enjoyed by the invariant tensors $\lambda$'s and $\eta$'s (the corresponding gauge parameters are ``hidden'' in these tensors).\\\\Also, in \cite{Capri:2018gpu} a gauge invariant mass term is introduced. This invariant mass term is constructed by means of a gauge invariant composite superfield, $\mathbf{V}(V,\Xi,\bar{\Xi})$, given by,
\begin{equation}
\exp\Big[\,\mathbf{V}(V,\Xi,\bar{\Xi})\,\Big]=e^{-i\bar{\Xi}}e^{V}e^{i\Xi}\,,
\end{equation}
where $V=V^{A}T^{A}$ is the usual vector superfield and $\{\Xi,\bar{\Xi}\}$ pair of chiral conjugated Stueckelberg-like  superfields. Being dimensionless, the invariant composite field $\mathbf{V}$ gives rise to the the following mass term:
\begin{equation}
S_{m^{2}}=m^{2}\,\int dV\,\left(\mathbf{V}^{A}\mathbf{V}^{A}+t^{ABCD}\,\mathbf{V}^{A}\mathbf{V}^{B}\mathbf{V}^{C}\mathbf{V}^{D}
+t^{ABCDE}\,\mathbf{V}^{A}\mathbf{V}^{B}\mathbf{V}^{C}\mathbf{V}^{D}\mathbf{V}^{E}+\dots\right)\,,
\end{equation}
where $m^{2}$ is a mass squared parameter and $t^{ABCD\dots}$ invariant tensors. It is immediate to observe that there is a mass degeneracy among the $(n^{2}-1)$ directions of the group. Therefore, once we have at our disposal the GSMAG, which naturally splits the diagonal and off diagonal sectors of the group, it is possible to partially breaks the mass degeneracy and define two different mass parameters, one for the $(n-1)$ diagonal components and the other one for the $n(n-1)$ off-diagonal components. This approach is already being developed in the context of the ordinary Yang-Mills \cite{InProgress}, opening a way for the study of the so-called Abelian dominance conjecture.\\\\Another problem that can be investigated in the GSMAG is the Gribov problem \cite{Gribov:1977wm,Vandersickel:2012tz}. In the Landau gauge this problem was firstly investigated in superspace in Ref.~\cite{Amaral:2013uya}. In non-supersymmetric scenario the Gribov ambiguity problem was extensively investigated in the maximal Abelian gauge in Refs \cite{Capri:2005tj,Capri:2006cz,Capri:2008ak,Capri:2013vka,Capri:2010an}.

\section*{Acknowledgments}
The Conselho Nacional de Desenvolvimento Cient\'{\i}fico e
Tecnol\'{o}gico (CNPq-Brazil), the
Coordena{\c{c}}{\~{a}}o de Aperfei{\c{c}}oamento de Pessoal de
N{\'{\i}}vel Superior (CAPES) and the SR2-UERJ are gratefully acknowledged. M. A. L. Capri is a level PQ-2 researcher under
the program Produtividade em Pesquisa-CNPq, grant n. $307783/2014-6$ and is a Procientista under SR2-UERJ. R. C. Terin is supported by the Coordena{\c{c}}{\~{a}}o de Aperfei{\c{c}}oamento de Pessoal de
N{\'{\i}}vel Superior (CAPES) under the program Doutorado Sandu\'{i}che no Exterior (PDSE), grant n. $88881.188419/2018-01$. 


\appendix

\section{The maximal Abelian gauge for ordinary SU($n$) Yang-Mills theory}
\label{MAGYM}
In this section, we make a brief review on the ordinary maximal Abelian gauge. This gauge arises from the breaking of color symmetry, which generates a separation of the structure of the group SU($n$). It is well-known that the gauge field is defined in the adjoint representation of SU($n$) group, namely 
\begin{equation}
A_{\mu}(x) =A_{\mu}^{A}(x)T^{A}\,, 
\end{equation}
where, the index $A$ runs from $1$ to $(n^{2}-1)$ and $T^{A}$ stands for a set of Hermitean traceless matrices forming the generators of the group. These generators can be split into a diagonal sector and a off-diagonal sectors, $T^{A} \equiv \{T^{a},T^{i}\}$, where the indices $\{a,b,c,...\} = 1,...,n(n-1)$ are the so-called off-diagonal indices, connected to the non-abelian sector, while the indices $\{i,j,k,...\} = 1,...,n-1$ are the diagonal ones, related to the Abelian subgroup of SU($n$). In this way, the Lie algebra,
\begin{equation}
\Big[T^{A},T^{B}\Big]=if^{ABC}T^{C}\,,
\end{equation}
is rewritten in terms of the diagonal and off-diagonal components, i.e.
\begin{eqnarray}
\Big[T^{a},T^{b}\Big] &=& if^{abc}T^{c} + if^{abi}T^{i}\,,\\
\Big[T^{a},T^{i}\Big] &=& -if^{abi}T^{c}\,,\\
\Big[T^{i},T^{j}\Big]&=&0\,,
\end{eqnarray}
where $f^{abc}$ e $f^{abi}$ are the structure constants of the group. These constants obey the following Jacobi's relations
\begin{eqnarray}
0 &=& f^{abi}f^{bcj} + f^{abj}f^{bic}\,,\\
0 &=& f^{abc}f^{cdi} + f^{adc}f^{cib} + f^{aic}f^{cbd}\,,\\
0 &=& f^{abc}f^{cde} + f^{abi}f^{ide} + f^{adc}f^{ced} + f^{abi}f^{ieb} + f^{aec}f^{cbd} + f^{aei}f^{ibd}\,,
\end{eqnarray}
which are derived from
\begin{equation}
f^{ABC}f^{CDE}+f^{ADC}f^{CEB}+f^{AEC}f^{CBD}=0\,.
\end{equation}
Now the gauge field can be split in terms of the two components of the group,
\begin{equation}
A_{\mu}(x) =A_{\mu}^{A}(x)T^{A} = A_{\mu}^{a}(x)T^{a} + A_{\mu}^{i}(x)T^{i}.  
\end{equation}
Then, the Yang-Mills action can also be written in terms of its Abelian and non-Abelian components,
\begin{equation}
S_{\mathrm{YM}} = -\frac{1}{2g^{2}}\mathrm{Tr}\int d^{4}x\, F^{\mu\nu}F_{\mu\nu} = -\frac{1}{4g^{2}}\int d^{4}x\, \left(F^{a \mu\nu}F^{a}_{\mu\nu} + F^{i \mu\nu}F^{i}_{\mu\nu}\right)
\end{equation}
where,
\begin{equation}
\mathrm{Tr}\Big(T^{A}T^{B}\Big)=\frac{\delta^{AB}}{2}
\end{equation}
and
\begin{eqnarray}
F_{\mu\nu}&=&F^{A}T^{A}\,\,=\,\,F^{a}_{\mu\nu}T^{a}+F^{i}_{\mu\nu}T^{i}\,,\\
F^{a}_{\mu\nu}&=&D^{ab}_{\mu}A^{b}_{\nu}-D^{ab}_{\nu}A^{b}_{\mu}+f^{abc}A^{a}_{\mu}A^{b}_{\nu}\,,\\
F^{i}_{\mu\nu}&=&\partial_{\mu}A^{i}_{\nu}-\partial_{\nu}A^{i}_{\mu}+f^{abi}A^{a}_{\mu}A^{b}_{\nu}\,,\\
D^{ab}_{\mu}&=&\delta^{ab}\partial_{\mu}-f^{abi}A^{i}_{\mu}\,.
\end{eqnarray}
Moreover, this action is invariant by the following infinitesimal transformations,
\begin{eqnarray}
A^{i}_{\mu}\to(A^{\omega})^{i}_{\mu}&=&A^{i}_{\mu}-\left(\partial_{\mu}\omega^{i}+f^{abi}A^{a}_{\mu}\omega^{b}\right)\,,\\
A^{a}_{\mu}\to(A^{\omega})^{a}_{\mu}&=&A^{a}_{\mu}-\left(D^{ab}_{\mu}\omega^{b}+f^{abc}A^{b}_{\mu}\omega^{c}+f^{abi}A^{b}_{\mu}\omega^{i}\right)\,,
\label{gauge}
\end{eqnarray}
where the infinitesimal  gauge parameter, $\omega$, can also be split in terms of the two sectors of the group:
\begin{equation}
\omega(x)=\omega^{i}(x)T^{i}+\omega^{a}(x)T^{a}.
\end{equation}
Since the Yang-Mills action can divided from the point of view of the group structure, let us now verify which are the consequences arising from the quantization process of this theory. Let us start by defining the generator functional for the Green functions of the theory,
\begin{equation}
Z[J] \propto \int [dA]\,\exp\left(iS_{\mathrm{YM}} + i\int d^{4}x J^{A\mu}A_{\mu}^{A}\right)\,.
\end{equation}
However, since gauge symmetry is preserved, the functional integration measure, $[dA]$, overcounts equivalent configurations in the field space and therefore an additional condition, or constraint, must be imposed so that spurious degrees of freedom be eliminated. This constraint is the so-called \textit{gauge-fixing condition} and, according to the Faddeev-Popov quantization approach \cite{Bailin:1986wt}, it is introduced in the functional integrals as follows,
\begin{equation}
Z[J] \propto \int [dA]\,\delta(G)\det\left|\frac{\delta G[A^{\omega}]}{\delta \omega}\right|\,\exp\left(iS_{\mathrm{YM}} + i\int d^{4}x J^{A\mu}A_{\mu}^{A}\right)\,.
\label{Z_functional}
\end{equation}
Here, the $G$ functional plays the role of setting the necessary conditions to correct the problem of functional integration measure. The interesting point here, is that this functional can be choosen in different ways for the diagonal, $G^{i}$, and off-diagonal, $G^{a}$, sectors, i.e. is precisely here that the color symmetry breaking takes place. In the case of the maximal Abelian gauge, the off-diagonal condition can be obtained through the extremization of the following auxiliary functional\footnote{To impose the minimum condition it is necessary to take into account also $\delta^{2}H[A]>0$. This implies that the Faddeev-Popov operator, $\frac{\delta G[A^{\omega}]}{\delta \omega}$, must be positive definite. This extra condition also implies that the domain of functional integral is restricted to a region called Gribov's region and its implementation is well-defined only in Euclidean space \cite{Gribov:1977wm,Vandersickel:2012tz}.}$^{\mathrm{,}}$\footnote{The Landau gauge can also be defined as an extreme condition of an auxiliary functional. In this case, the suitable functional is $\int d^{4}x\,A^{A}_{\mu}A^{A\mu}$.},
\begin{equation}
H[A] = \frac{1}{2}\int d^{4}x A^{a}_{\mu}A^{a\mu}\,.
\end{equation}
Applying the extreme condition,
\begin{equation}
\delta H[A] = 0\,,
\end{equation}
we have then,
\begin{equation}
G^{a}[A]= D_{\mu}^{ab}A^{b\mu} = 0\,.
\label{MAG10}
\end{equation}
Since the symmetry subgroup $U(1)^{N-1}$ is present in the theory, it is also necessary to choose a gauge condition for the diagonal components. For simplicity, a Landau-like gauge condition is taken,
\begin{equation}
G^{i}[A] = \partial_{\mu}A^{i\mu} = 0\,.
\label{MAG20}
\end{equation} 
Thus, the set of equations \eqref{MAG10} and \eqref{MAG20} forms the so-called maximal Abelian gauge. It is important to mention here that the diagonal condition do not follows from a extreme condition of any auxiliary functional and different choices could be taken for this sector. This is the case, for example, of the modified maximal Abelian gauge \cite{Dudal:2005zr}.



\end{document}